\documentclass[apj,twocolumn]{emulateapj}
\usepackage{apjfonts}

\newcommand{\beq}{\begin{equation}}
\newcommand{\eeq}{\end{equation}}
\newcommand{\beqn}{\begin{eqnarray}}
\newcommand{\eeqn}{\end{eqnarray}}
\newcommand{\pd}{\partial}

\newcommand{\bx}{{\bf x}}
\newcommand{\bX}{{\bf X}}
\newcommand{\AU}{{\rm AU}} 

\newcommand{\kB}{k_{\rm B}}
\newcommand{\ovl}[1]{ {\overline{#1}} }

\newcommand{\bracket}[1]{\langle #1 \rangle}
\newcommand{\eqref}[1]{(\ref{#1})}
\newcommand{\dfrac}[2]{ {\displaystyle\frac{#1}{#2}} }

\newcommand{\pfrac}[2]{ \Bigl(\dfrac{#1}{#2}\Bigr) }
\newcommand{\eps}{\epsilon}
\renewcommand{\leq}{\leqslant}
\renewcommand{\geq}{\geqslant}

\defcitealias{O09}{O09}
\defcitealias{OTS09}{OTS09}
\defcitealias{OTTS11b}{Paper II}

\shorttitle{ELECTROSTATIC BARRIER AGAINST DUST GROWTH. I}
\shortauthors{OKUZUMI ET AL.}
\slugcomment{ApJ Accepted}

\begin{document}
\title{Electrostatic Barrier Against Dust Growth in Protoplanetary Disks. \\I. 
Classifying the Evolution of Size Distribution}
\author{Satoshi Okuzumi\altaffilmark{1,2}, Hidekazu Tanaka\altaffilmark{3},
Taku Takeuchi\altaffilmark{3,4}, and Masa-aki Sakagami\altaffilmark{1}}
\email{okuzumi@nagoya-u.jp}
\altaffiltext{1}{Graduate School of Human and Environmental Studies, Kyoto University,
Kyoto 606-8501, Japan}
\altaffiltext{2}{Department of Physics, Nagoya University, Nagoya, Aichi 464-8602, Japan}
\altaffiltext{3}{Institute of Low Temperature Science, Hokkaido University, Sapporo 060-0819, Japan}
\altaffiltext{4}{Department of Earth and Planetary Sciences, Tokyo Institute of Technology, Tokyo 152-8551, Japan}

\begin{abstract}
Collisional growth of submicron-sized dust grains into macroscopic aggregates
is the first step of planet formation in protoplanetary disks.
These grains are expected to carry nonzero negative charges
in the weakly ionized disks, 
but its effect on their collisional growth has not been fully understood so far.  
In this paper, we investigate how the charging affects the evolution of the dust size distribution 
properly taking into account the charging mechanism in a weakly ionized gas
as well as porosity evolution through low-energy collisions.
To clarify the role of the size distribution, we divide our analysis into two steps. 
First, we analyze the collisional growth of charged aggregates 
assuming a monodisperse (i.e., narrow) size distribution. 
We show that the monodisperse growth stalls due to the electrostatic repulsion 
when a certain condition is met, as is already expected in the previous work. 
Second, we numerically simulate dust coagulation using Smoluchowski's method 
to see how the outcome changes when the size distribution is allowed to freely evolve.
We find that, under certain conditions, the dust undergoes bimodal growth
where only a limited number of aggregates continue to grow carrying the major part 
of the dust mass in the system.
This occurs because remaining small aggregates efficiently sweep up free electrons 
to prevent the larger aggregates from being strongly charged.
We obtain a set of simple criteria that allows us to predict how the size distribution evolves 
for a given condition.
In Paper II, we apply these criteria to dust growth in protoplanetary disks.
\end{abstract}
\maketitle

\section{Introduction}
The standard core-accretion scenario for planet formation
\citep{Mizuno80,Pollack+96} is based on the so-called planetesimal hypothesis.
This hypothesis assumes that solid bodies of 
size larger than kilometers (called ``planetesimals'') form in a protoplanetary disk prior to planet formation.
However, the typical size of solid particles in interstellar space 
is as small as a micron or even smaller \citep{MRN77}.
It is still open how the submicron-sized grains 
evolved into kilometer-sized planetesimals. 

The simplest picture for dust evolution towards planetesimals can be summarized 
into the following steps.
(1) Initially, submicron-sized particles coagulate into larger but highly porous, fractal aggregates 
through low-velocity collisions driven by 
Brownian motion and differential settling towards the midplane of the disk 
\citep{WB98,Blum+98,KPH99}.
(2)  As the aggregates grow to ``macroscopic'' (mm to cm) sizes, 
the collisional energy becomes high enough to cause the compaction of the aggregates 
\citep{Blum04,SWT08,PD09}.
(3)The compaction cause the increase in the stopping times of the aggregates,
allowing them to concentrate in the midplane of the disk \citep{Safronov69,GW73},
the center of vortices \citep{BS95}, or turbulent eddies \citep{J+07}.
(4) Planetesimals may form within such dense regions through gravitational instability 
\citep{Safronov69,GW73} or through further collisional growth \citep{WC93,W95}.

However, there is great uncertainty on how large dust aggregates can grow 
through mutual collisions \citep[see, e.g.,][]{BW08,Guettler+10}.
As the collisional compaction proceeds, the aggregates decouple from the ambient gas
and obtain higher and higher relative velocities driven by radial drift 
\citep{W77} and gas turbulence \citep{Voelk+80}.
The collision velocity can exceed $10~{\rm m~s^{-1}}$ even without turbulence,
but it is uncertain whether such high-speed collisions lead to the sticking 
or fragmentation of the aggregates 
\citep{BW08, Wada+09,TW09,Guettler+10}.
In addition, collisional compaction itself can cause the reduction of sticking efficiency 
\citep{BW08,Guettler+10}.
This may terminates the collisional growth before the fragmentation occurs  \citep{Zsom+10}.

By contrast, it is generally believed that dust coagulation proceeds rapidly 
until the aggregates grow beyond the initial, fractal growth stage
since the collision velocity is too low to cause the reduction of sticking efficiency 
\citep{DT97,BW08,Guettler+10}.
However, one of the authors has recently pointed out that electric charging of aggregates 
could halt dust growth {\it before} the aggregates leave this stage \citep[][hereafter O09]{O09}.
Protoplanetary disks are expected to be weakly ionized 
by a various kinds of high-energy sources, such as cosmic rays \citep{UN81}
and X-rays from the central star \citep{GNI97}.
In such an ionized environment, 
dust particles charge up by capturing ions and electrons,
as is well known in plasma physics \citep{SM02}.
In equilibrium, dust particles acquire nonzero {\it negative} net charges
because electrons have higher thermal velocities than ions.
This ``asymmetric'' charging causes a repulsive force between colliding aggregates,
but this effect has been ignored in previous studies on protoplanetary dust growth.
\citetalias{O09} has found that the dust charge in a weakly ionized disk can be considerably smaller
than in a fully ionized plasma but can nevertheless inhibit dust coagulation in a wide region of the disk.
It is also found that the electrostatic barrier becomes significant 
when the dust grows into fractal aggregates,
 i.e., much earlier than the growth barriers mentioned above emerge.
Thus, the dust charging can greatly modify the current picture of 
dust evolution towards planetesimals.

The analysis of the electrostatic barrier by \citetalias{O09} 
is based on the assumption that dust aggregates obey a narrow size distribution. 
In reality, however, size distribution is determined as a result of the coagulation process, 
and it has been unclear how the distribution evolves when the dust charging is present.
The purpose of this study is to clarify how the size distribution of dust aggregates 
evolves when the aggregates are charged in a weakly ionized gas.

According to \citetalias{O09}, the effect of dust charging can become already significant 
before the collisional compaction of aggregates becomes effective.
In this stage, dust aggregates are expected to 
have lower and lower internal density (i.e., higher and higher porosity) as they grow,
as is suggested by laboratory experiments and $N$--body simulations
\citep{WB98,Blum+98,KPH99}.
This porosity evolution has been ignored in most theoretical studies on dust coagulation
\citep[e.g.,][]{NNH81,THI05,BDH08}, in which aggregates are simplified as compact spheres.
However, when analyzing the electrostatic barrier, 
the porosity evolution must be accurately taken into account;
in fact, as we will see later, the ignorance of the porosity evolution 
leads to considerable underestimation of the electrostatic barrier,
because compact spheres are generally less coupled to the ambient gas
and hence have higher collision energies than porous aggregates.
In this study, we use the fractal dust model recently proposed by
\citet[][hereafter \citetalias{OTS09}]{OTS09}.
Classically, fractal dust growth has been only modeled with either of its two extreme limits,
namely, ballistic cluster-cluster and particle-cluster aggregation
\citep[BCCA and BPCA; e.g.,][]{Ossenkopf93,DD05}.
To fill the gap between the two limits, 
\citetalias{OTS09} introduced a new aggregation model 
(called the quasi-BCCA model) in which aggregates grow through unequal-sized collisions.
\citetalias{OTS09} found from $N$--body simulations 
that the resultant aggregates tend to have a fractal dimension $D$ close to $2$
even if the size ratio deviates from unity. 
This explains why fractal aggregates with $D\sim 2$ are 
universally observed in various low-velocity coagulation processes 
\citep{WB98,Blum+98,KPH99}.
\citetalias{OTS09} summarized the results of their $N$--body simulations 
into a simple analytic formula giving the increase in the porosity (volume) 
for general hit-and-stick collisions.
This formula together with the Smoluchowski equation extended
for porous dust coagulation \citepalias{OTS09} 
enables us to follow the evolution of size distribution and porosity
consistently with dust charging.

As we will see later, our problem involves many model parameters,
such as the initial grain size and the gas ionization rate.
To fully understand the dependence of the results on these parameters,
we do not assume any protoplanetary disk model 
but seek to find general criteria determining the outcome of dust evolution.
This approach allows us to investigate the effect of the electrostatic barrier
with any protoplanetary disk models.
Application of the growth criteria to particular disk models will be done 
in \citetalias{OTTS11b} \citep{OTTS11b}.

This paper is structured as follows.
In Section~2, we describe the dust growth model used in this study.
In Section~3, we examine the case of monodisperse growth 
in which all the aggregates grow into equal-sized ones.
The monodisperse model allows us to introduce several important quantities governing 
the outcome of the growth.
We analytically derive a criterion in which the ``freezeout'' of monodisperse growth occurs. 
In Section~4, we present numerical simulations including the evolution 
of the size distribution to show how the outcome of the growth differs from 
the prediction of the monodisperse theory.
We discuss the validity of our dust growth model in Section~5.
A summary of this paper is presented in Section~6.
 
\section{Dust Growth Model}
In this section, we describe the dust growth model considered in this study.

\begin{figure}
\centering
\plotone{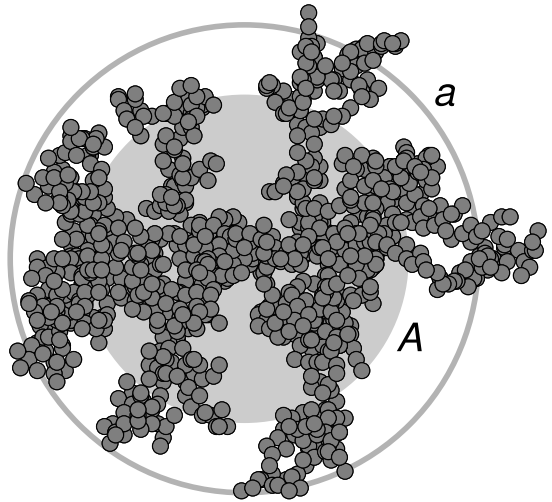}
\caption{
Projection of a numerically created, three-dimensional porous aggregate
consisting of $\approx 1000$ monomers. 
The large open circle shows the characteristic radius $a$ (for its definition, see Section 2.3.1),
while the gray disk inside the circle shows the projected area $A$ 
averaged over various projection angles. 
Note that $A$ is not necessarily equal to $\pi a^2$, 
especially when the aggregate is highly porous (see also Figure~4 of \citetalias{OTS09}).
}
\label{fig:Nbody}
\end{figure}
We consider collisional growth of dust starting from an ensemble of 
equal-sized spherical grains (``monomers'').
Each aggregate is characterized 
by its mass, radius, projected area, and charge.
For simplicity, we assume ``local'' growth, i.e., we neglect global transport of dust within a disk.

We focus on the first stage of dust evolution in protoplanetary disks 
and assume that aggregates grow through ``hit-and-stick'' collisions,
i.e., collisions with perfect sticking efficiency and no compaction.
It is known theoretically \citep[e.g.][]{KPH99} and experimentally \citep[e.g.][]{WB98} that 
hit-and-stick collisions lead to highly porous aggregates.
To take into account the porosity evolution,
we adopt the fractal dust model proposed by \citetalias{OTS09}.
This model characterizes each aggregate with its mass $M$ and ``characteristic radius'' $a$
(see \citetalias{OTS09} and Section~2.3 for the definition of the characteristic radius),
and treat the two quantities as independent parameters.
Another important parameter is the projected area $A$
This determines how the aggregates are frictionally coupled to the gas.
In the \citetalias{OTS09} model, $A$ is not treated as an independent parameter but
 is given as a function of $M$ and $a$. 
Note that $A$ is not generally equal to a naive ``cross section'' $\pi a^2$, 
especially when the aggregates is highly porous (Figure~\ref{fig:Nbody}; 
see also Figure 4 of \citetalias{OTS09}).
Distinction between $A$ and $\pi a^2$ allows us 
to avoid overestimation of the gas drag force to dust aggregates.  
In Section~2.3, We will describe the porosity model in more detail. 

The collision probability between two aggregates 1 and 2 is proportional 
to their relative speed $\Delta u$ times the collisional cross section $\sigma_{\rm coll}$
given by \citep[e.g.,][]{LL76}
\beq
\sigma_{\rm coll} =  \left\{ \begin{array}{ll}
\pi (a_1+a_2)^2\left(1 - \dfrac{E_{\rm el}}{E_{\rm kin}} \right), & E_{\rm kin}>E_{\rm el}  \\[3pt]
0,  &  E_{\rm kin} \leq E_{\rm el}, 
\end{array} \right.
\label{eq:sigma_coll}
\eeq
where $E_{\rm kin} = M_\mu(\Delta u)^2/2$ is the kinetic energy associated with the relative motion,
$M_\mu = M_1M_2/(M_1+M_2)$ is the reduced mass, and
$E_{\rm el}= Q_1 Q_2/(a_1 + a_2)$ is the energy needed for the aggregates to collide with each other.
In this paper, $E_{\rm el}$ is called ``the electrostatic energy'' for colliding aggregates.
Below, we describe how to determine $Q$ and $\Delta u$.

\subsection{Charging}
We adopt the dust charging model developed by \citetalias{O09}.
In this model, dust aggregates are surrounded by a weakly ionized gas 
and charge up by capturing free electrons and ions. 
These ionized particles are created by the nonthermal ionization of the neutral gas
and are removed from the gas phase through the adsorption to the dust 
as well as the gas-phase recombination.
The dust charge $Q$ and the number densities of ions and electrons 
are thus determined by the balance among the ionization, recombination, and dust charging. 
In equilibrium, the average charge $\bracket{Q}_a$ of aggregates with radius $a$  is given by
(see Equation~(23) of \citetalias{O09})
\beq
\bracket{Q}_a = -\Psi\frac{a \kB T}{e},
\label{eq:Q}
\eeq
where $\kB$ is the Boltzmann constant, $T$ is the gas temperature,
$e$ is the elementary charge,
and $\Psi$ is a dimensionless parameter characterizing the charge state of the gas-dust mixture.
\citetalias{O09} has analytically shown that the equilibrium conditions
are reduced to a single equation for $\Psi$.
When the adsorption to the dust dominates the removal of the ionized gas,
the equation for $\Psi$ is written as (see Equation~(34) of \citetalias{O09})

\beq
\frac{1}{1+\Psi} - \frac{s_i}{s_e}\sqrt{\frac{m_e}{m_i}}\exp\Psi - \frac{\Psi}{\Theta} = 0,
\label{eq:Psi}
\eeq
where $m_{i(e)}$ is the mass of ions (electrons), 
$s_{i(e)}$ is their sticking probability onto a dust monomer, and 
\beq
\Theta = \frac{\zeta n_g e^2}{s_i  A_{\rm tot}C_{\rm tot}\kB T} \sqrt\frac{\pi m_i}{8\kB T},
\label{eq:Theta_def}
\eeq 
is a dimensionless quantity depending on 
the total projected area $A_{\rm tot} = \int A(M) n(M) dM$ and
total radius $C_{\rm tot} = \int a(M) n(M) dM$ of aggregates, 
and the ionization rate $\zeta$ and number density $n_g$ of neutral gas particles.
Equation~\eqref{eq:Psi} originates from the quasi-neutrality condition,
$en_i - en_e + Q_{\rm tot} = 0$, where $n_i$ and $n_e$ are the 
number density of ions and electrons, and 
$Q_{\rm tot} = \int \bracket{Q}_{a(M)} n(M)dM$ is the total charge
carried by dust in a unit volume.\footnote{
$n_i$ and $n_e$ are related to $\Psi$ as \citepalias{O09}
\beqn
n_i = \frac{\zeta n_g}{s_i A_{\rm tot}} \sqrt\frac{\pi m_i}{8\kB T}\frac{1}{1+\Psi}, \quad 
n_e = \frac{\zeta n_g}{s_e A_{\rm tot}}  \sqrt\frac{\pi m_e}{8\kB T}\exp\Psi, \nonumber
\eeqn
}
Equation~\eqref{eq:Psi} cannot be used when the gas-phase recombination 
dominates the removal of the ionized gas.
In a typical protoplanetary disk, however, the gas-phase recombination can be safely neglected 
unless the dust-to-gas ratio is many orders of magnitude 
smaller than interstellar values $\sim 0.01$ \citepalias{O09}.

Physically, $\Psi$ is related to the surface potential of aggregates.
For an aggregate with charge $Q$ and radius $a$, the surface potential $\psi$ is given by $\psi = Q/a$.
Equation~\eqref{eq:Q} implies that $\Psi = \bracket{\psi}_a/(-\kB T/e)$, namely, 
$\Psi$ is the surface potential averaged over aggregates of radius $a$ and normalized by $-\kB T/e$.
Note that $\bracket{\psi}_a$ is apparently independent of $a$, but is actually not
because $\Psi$ depends on the size distribution of aggregates through $A_{\rm tot}$ and $C_{\rm tot}$.
It should be also noted that the radius $a$ 
can be interpreted as the {\it electric capacitance} $C$ (i.e., $Q = C\psi$).
This is the reason why we have denoted the total radius as $C_{\rm tot}$.

\begin{figure}
\plotone{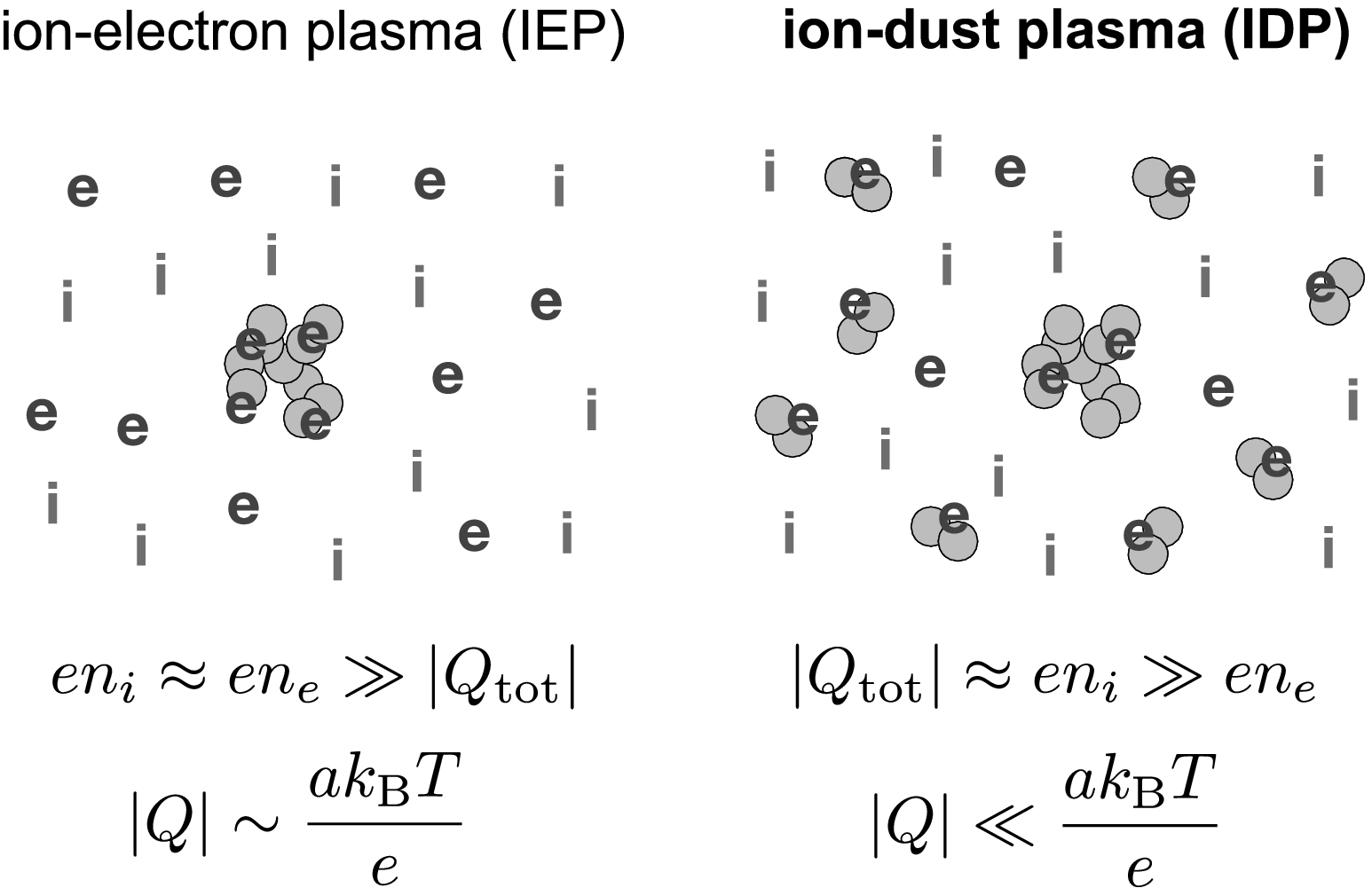}
\caption{
Schematic illustration of an ion-electron plasma (IEP: left) and an ion-dust plasma (IDP; right).
In an IEP, the dominant carriers of negative charges are free electrons.
In an IDP, by contrast, the dominant negative species is the charged dust.
The absolute value of the dust surface potential, $|\psi|=a|Q|$, 
is generally smaller in IDPs than in IEPs. 
}
\label{fig:plasma}
\end{figure}

As shown in \citetalias{O09}, $\Psi$ asymptotically behaves as
(see Section~2.3 of \citetalias{O09})
\beq
\Psi \approx  \left\{ \begin{array}{ll}
\Psi_\infty, & \Theta \gg \Psi_\infty,   \\[3pt]
\Theta, & \Theta \ll \Psi_\infty,
\end{array} \right.
\label{eq:Psi_asympt}
\eeq
where $\Psi_\infty$ is the solution to
\beq
\frac{1}{1+\Psi_\infty} - \frac{s_i}{s_e}\sqrt{\frac{m_e}{m_i}}\exp\Psi_\infty = 0.
\label{eq:Psi_infty}
\eeq
Equation~\eqref{eq:Psi_infty} is known as the equation for the equilibrium charge  of a dust particle 
embedded in a fully ionized plasma \citep{Spitzer41,SM02}.
Equation~\eqref{eq:Psi_asympt} suggests that the charge state of dust particles
in a weakly ionized gas is characterized by two limiting cases.
If $\Theta \gg \Psi_\infty$, the total negative charge $|Q_{\rm tot}|$
carried by dust aggregates is negligibly small compared to $e n_e$,
and the quasi-neutrality condition approximately hold in the gas phase, i.e., $n_i \approx n_e$.
If $\Theta \ll \Psi_\infty$, by contrast, most of the negative charge in the system
is carried by aggregates, and the quasi-neutrality condition 
approximately holds between ions and negatively charged dust.
For this reason, \citetalias{O09} referred to the former phase as the {\it ion-electron plasma} (IEP), 
and to the latter as the {\it ion-dust plasma} (IDP). 
Figure~\ref{fig:plasma} schematically shows the difference between the two plasma states.

For given $m_{i}$ and $s_i/s_e$, 
Equation~\eqref{eq:Psi} determines $\Psi$ as a function of $\Theta$.
In typical protoplanetary disks, the dominant ion species are molecular ions  (e.g., ${\rm HCO}^+$)
or metal ions (e.g., ${\rm Mg}^+$) depending on the abundance of metal atoms
in the gas phase \citep{Sano+00,IN06}.
Although $s_i$ is likely to be close to unity \citep{UN80,DS87}, 
$s_e$ at low temperatures is poorly understood.
\citet{Umebayashi83} estimated $s_e$ using a semiclassical phonon theory to obtain $0.1 \la s_e \la 1$
for $T \la 100{\rm K}$.
However, the uncertainty in $s_e$ does not strongly affect the evaluation of $\Psi$.
For example, assuming $m_i = 24m_{\rm H}$ (the mass of Mg$^+$) and $s_i = 1$, 
$\Psi_\infty$ is 3.78 for $s_e = 1$, and is $1.96$ even for $s_e = 0.1$. 

\begin{figure}
\centering
\plotone{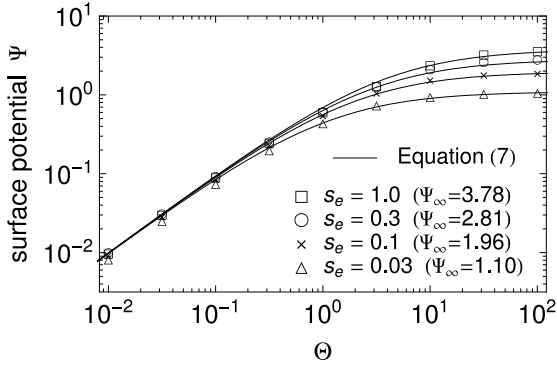}
\caption{
Comparison between the numerical solutions to Equation~\eqref{eq:Psi} 
and the approximate formula~\eqref{eq:Psi_approx}. 
The symbols indicate the numerical solutions for various values of $s_e$, 
and the solid curves show the prediction from Equation~\eqref{eq:Psi_approx}.
The ion mass is taken to be $24m_{\rm H}$ for all the cases. 
The maximum values $\Psi_\infty$ are $3.78$, $2.81$, $1.96$, and $1.10$ for
$s_e=$ 1.0, 0.3, 0.1, 0.03, respectively.
}
\label{fig:Psi}
\end{figure}

Figure~\ref{fig:Psi} illustrates the dependence of $\Psi$ on $\Theta$ 
for fixed $m_i (= 24m_{\rm H})$ and $s_i(=1)$ with various $s_e$(=1.0, 0.3, 0.1, 0.03).
We find that $\Psi$ can be well approximated by
\beq
\Psi \approx \Psi_\infty \left[1+ \pfrac{\Theta}{\Psi_\infty}^{-0.8} \right]^{-1/0.8}.
\label{eq:Psi_approx}
\eeq
In Figure~\ref{fig:Psi}, we compare Equation~\eqref{eq:Psi_approx} 
with the numerical solutions to the original equation.
The approximate formula recovers all the numerical solutions within an error of $20\%$.
This means that $\Psi/\Psi_\infty$ is well approximated as a function of $\Theta/\Psi_\infty$
for this parameter range.\footnote{
This is not true for more general cases. 
In fact, Equations~\eqref{eq:Psi} and \eqref{eq:Psi_infty} can be combined into
a single equation (Equation~\eqref{eq:Psi2}),
which cannot be reduced to an equation for $\Psi/\Psi_\infty$
depending only on $\Theta/\Psi_\infty$.
}
We use this fact in Section 3.

Up to here, we have considered only the mean value of the charge $Q$.
In fact,  there always exists a finite value of the charge dispersion $\bracket{\delta Q^2}_a$,
and moreover, the mean value $\bracket{Q}_a$ is not necessarily larger than 
$\bracket{\delta Q^2}^{1/2}_a$ \citepalias{O09}.
Nevertheless, we will assume below that  
{\it the dust charge $Q$ is always equal to $\bracket{Q}_a$}.
The validity of this assumption will be discussed in Section~5.2.
 
\subsection{Dust Dynamics}
As found from Equation~\eqref{eq:sigma_coll}, 
the relative velocity between aggregates determines whether they 
can overcome the electrostatic barrier to collide. 
In this study, we model the motion of dust aggregates in the following way.
We assume that the motion of each aggregate relative to the ambient gas 
consists of random Brownian motion and systematic drift due to spatially uniform acceleration
(e.g., uniform gravity).
Then, the probability density function $P_r(\Delta{\bf u})$ 
for the relative velocity $\Delta {\bf u} \equiv {\bf u}_1 -  {\bf u}_2  $ 
between two aggregates 1 and 2 is given by
\beq
P_r(\Delta{\bf u}) d \Delta {\bf u} = \pfrac{M_\mu}{2\pi \kB T}^{3/2} 
\exp \left( -\frac{M_\mu(\Delta{\bf u}-\Delta{\bf u}_D)^2}{2\kB T}\right) d \Delta {\bf u},
\label{eq:PDF} 
\eeq
where $\Delta {\bf u}_D$ is the difference of the drift velocities between the two aggregates.
Here, we have assumed that the systematic motion has no fluctuating component,
that is, the velocity dispersion is thermal even when $M_\mu \Delta u_D \gg k_{\rm B}T$.
We will discuss the effect of adopting a different velocity distribution in Section~5.3. 

We further assume that aggregates are frictionally coupled to the ambient gas,
and give $\Delta {\bf u}_D$ as
\beq
\Delta {\bf u}_D = {\bf g}(\tau_1-\tau_2),
\label{eq:uD}
\eeq
where $\tau_j (j=1,2)$ is the stopping time of each aggregate and ${\bf g}$ is the 
uniform acceleration.
In this study, we focus on small aggregates and give $\tau$ according to Epstein's law,
\beq
\tau = \frac{3M}{4\rho_gA}\sqrt{\frac{\pi m_g}{8 k_{\rm B}T}},
\label{eq:tstop}
\eeq
where $\rho_g$ is the gas density 
and $m_g$ is the mass of the gas particles.
Epstein's law is valid when the size $a$ of the aggregate is smaller the mean free path $\ell$ of gas particles. 

In a protoplanetary disk, relative motion like Equation~\eqref{eq:uD} is driven by several processes.
For example, the gravity of the central star causes acceleration $g = \Omega_{\rm K}^2 z$ towards the midplane of the disk, where $\Omega_{\rm K}$ 
is the Kepler rotational frequency and $z$ is the distance from the midplane.
Another example is the acceleration driven by gas turbulence in the strong coupling limit. 
When both of two colliding aggregates are frictionally well coupled to
the turbulent eddies of all scales,
the relative velocity between the aggregates is approximately given by
$\Delta u_D \approx ({u_\eta }/{t_\eta})|\tau_1 - \tau_2|$,  
where $u_\eta$ and $t_\eta$ are the characteristic velocity and turnover time 
for the smallest eddies, respectively \citep{W84,OC07}.
This means that turbulence behaves as an effective acceleration field 
of $g \approx  {u_\eta }/{t_\eta}$ for strongly coupled aggregates.

As the collisional cross section $\sigma_{\rm coll}$ depends on the stochastic variable $\Delta {\bf u}$,
it is useful to treat collision events statistically.
To do so, we introduce the collisional rate coefficient 
\beq
K \equiv \int P_r(\Delta {\bf u})\sigma_{\rm coll}|\Delta {\bf u}| d \Delta {\bf u}.
\eeq
With Equations~\eqref{eq:sigma_coll} and \eqref{eq:PDF}, 
the integration can be analytically performed.
Using $Q_1Q_2 >0$, we have \citep{Shull78}
\beqn 
K &=& \pi(a_1+a_2)^2\sqrt{\frac{\kB T}{2\pi M_\mu {\cal E}_D}}
\Bigl[
y_+ \exp(-y_-^2) - y_-  \exp(-y_+^2)  \nonumber \\
&& + \frac{\sqrt{\pi}}{2}(1-2y_+y_-)\left\{ {\rm erf}(y_+)- {\rm erf}(y_-)\right\}
\Bigr],
\label{eq:K}
\eeqn
where  ${\rm erf}(y) =(2/\sqrt{\pi})\int_0^y \exp(-z^2) dz $ is the error function, and
$y_+$ and $y_-$ are  defined as
\beq
y_\pm = \sqrt{{\cal E}_E} \pm \sqrt{{\cal E}_D}, 
\eeq
with
\beq
{\cal E}_D = \frac{M_\mu (\Delta u_D)^2}{2 \kB T},
\label{eq:ED_def}
\eeq
\beq
{\cal E}_E = \frac{Q_1 Q_2}{(a_1+ a_2)\kB T}.
\label{eq:EE_def}
\eeq
Note that ${\cal E}_D$ and ${\cal E}_E$ are the relative kinetic energy associated with differential drift 
and the electrostatic energy normalized by $\kB T$, respectively.

Equation~\eqref{eq:K} has the following simple asymptotic forms:
\beq
K \approx \left\{ \begin{array}{ll}
\pi(a_1+a_2)^2\Delta u_B \exp(-{\cal E}_E), & {\cal E}_D \ll 1,  \\[3pt]
\pi(a_1+a_2)^2\Delta u_D \left(1-\dfrac{{\cal E}_E}{{\cal E}_D}\right),
 &  {\cal E}_D \gg 1,{\cal E}_E, 
\end{array} \right.
\label{eq:K_approx}
\eeq 
where $\Delta u_B = (8\kB T/\pi M_\mu)^{1/2}$  is the mean thermal speed between the colliding aggregates.
The exponential factor $\exp(-{\cal E}_E)$ originates from the high-energy tail of the Maxwell distribution.
This factor guarantees $K$ nonvanishing even for large ${\cal E}_E$. 

\subsection{Porosity Model}
As shown by \citetalias{O09}, the charging affects
dust growth before the collisional compaction becomes effective.
In this early stage, aggregates have a highly porous structure \citep{WB98,KPH99}.  
The porosity influences their collisional growth 
through the collisional and aerodynamical cross sections.
It also affects dust charging through the capacity (=radius) 
and the capture cross section for ions and electrons.
Therefore, it is important to adopt a realistic model for the porosity of aggregates.

In this study, we adopt the porosity model developed by \citetalias{OTS09}.
This model is based on $N$--body simulations of successive collisions
between aggregates of various sizes.
This model provides a natural extension of the classical hit-and-stick aggregation models 
(see \citetalias{OTS09} and references therein).
Collisional fragmentation and restructuring is not taken into account, 
so the porosity increase only depends on the physical sizes of colliding aggregates.
This assumption is valid as long as the collisional energy is sufficiently lower than 
the critical energy for the onset of collisional compaction.
The validity of this assumption will be discussed in Section 5.4.

\subsubsection{Porosity Increase After Collision}
Our porosity model measures the size of a porous aggregate with the characteristic radius
$a \equiv [(5/3N) \sum_{k=1}^N (\bx_k - \bX)^2]^{1/2}$,
where $N$ is the number of constituent monomers within the aggregate,
$\bx_k$ is the coordinate of the $k$-th constituent monomer, and $\bX$ is the center of mass. 
Figure~\ref{fig:Nbody} shows the characteristic radius as well as the projected area $A$ 
of a numerically created porous aggregate.
In our model, the porosity of each aggregate is characterized by $a$ and $N$, 
while the projected area $A$ is assumed to be a function of them.
In the following subsections, we summarize how $a$ and $A$ are calculated in this model.

The porosity evolution of aggregates after a collision is expressed 
in terms of the increase in the porous volume $V \equiv  (4\pi/3)a^3$.    
For a collision between aggregates with volumes $V_1$ and $V_2 (\leq V_1)$,
the volume of the resulting aggregate, $V_{1+2}$, can be generally written as
\beq
V_{1+2} = V_{1}  + (1 + \chi )V_2,
\label{eq:V_1+2}
\eeq
where $\chi$ is a dimensionless factor depending on $V_1$ and $V_2$.
We refer to $\chi$ as the ``void factor'' since it identically vanishes for compact aggregation.

It is known that there are two limiting cases for hit-and-stick collisions
 \citep[see, e.g.,][]{Mukai+92,Kozasa+93}.
One is called the ballistic cluster--cluster aggregation (BCCA)
where aggregates grow only through equal-sized collisions.
On average, the characteristic radius of BCCA clusters is related to the monomer number $N$ as
\beq
a_{\rm BCCA} \approx a_0 N^{1/D_{\rm BCCA}},
\label{eq:a_BCCA}
\eeq
where $a_0$ is the radius of monomers and $D_{\rm BCCA} \approx 1.9$ 
is the fractal dimension of BCCA clusters \citep[e.g.,][]{Mukai+92}.
The void factor for the BCCA growth can be calculated from Equation~\eqref{eq:a_BCCA}
as $\chi_{\rm BCCA} = 2^{3/D_{\rm BCCA}}-2 \approx 0.99$ \citepalias{OTS09}.
The opposite limit is called the ballistic particle--cluster aggregation (BPCA), 
in which an aggregate grows by colliding with monomers.
On average, the characteristic radius of BPCA clusters is given by
$a_{\rm BPCA} \approx (1-P_{\rm BPCA})^{-1/3} a_0 N$, where
$P_{\rm BPCA} \equiv 1 - (Na_0/a_{\rm BPCA})^3 \approx 0.874$
 is the porosity of BPCA clusters \citep[e.g.,][]{Kozasa+93}.
The void factor is found to be $\chi_{\rm BPCA} = P_{\rm BPCA}/(1-P_{\rm BPCA}) \approx  6.94$
\citepalias{OTS09}. 
Note that both $\chi_{\rm BCCA}$ and $\chi_{\rm BPCA}$ are constant.
 
To obtain $\chi$ for more general cases, 
\citetalias{OTS09} presented a new aggregation model called the ``quasi-BCCA'' (QBCCA).
In the QBCCA, an aggregate grows through unequal-sized collisions with 
a fixed mass ratio $N_2/N_1$, where $N_1$ and $N_2 (<N_1)$ are the monomer numbers 
of the target and projectile, respectively.
The projectile is chosen among the outcomes of earlier collisions, so that 
the resultant aggregate has a self-similar structure.
\citetalias{OTS09} performed $N$-body simulations of aggregate collisions with various 
size ratios and found that the void factor for QBCCA is approximately given by
\beq
\chi_{\rm QBCCA}(V_1/V_2) = \chi_{\rm BCCA} - 1.03 \ln\pfrac{2}{V_{1}/V_{2} +1 }.
\label{eq:chi_QBCCA}
\eeq
Note that $\chi_{\rm QBCCA}$ approaches to $\chi_{\rm BCCA}$ in the BCCA limit
($V_1/V_2 \to 1$) as must be by the definition of BCCA.

Unfortunately, Equation~\eqref{eq:chi_QBCCA} does not reproduce the void factor 
in the BPCA limit ($V_1/V_2 \to \infty$).
To bridge the gap between the BCCA and BPCA limit, \citetalias{OTS09} considered a formula
\beq
\chi = \min \left\{ \chi_{\rm QBCCA}(V_1/V_2), \chi_{\rm BPCA} \right\}.
\label{eq:chi}
\eeq 
It is easy to check that Eqaution~\eqref{eq:chi} approaches to $\chi_{\rm BCCA}$ 
and $\chi_{\rm BPCA}$ in the BCCA and BPCA limits, respectively.
Equation~\eqref{eq:chi} will be used in the numerical simulations presented in Section~4
to determine the porosity (volume) of aggregates after collisions.

\subsubsection{Projected Area}
The projected area $A$ is another key property of porous aggregates.
This does not affect only the charge state of the gas-dust mixture
(Equation~\eqref{eq:Theta_def}) but also 
the drift velocity of individual aggregates (Equation~\eqref{eq:tstop}).

For BCCA clusters, the projected area 
averaged for fixed $N$ is well approximated by \citep{Minato+06} 
\beq
\ovl{A}_{\rm BCCA} = \pi a_0^2 \times \left\{ \begin{array}{ll}
12.5N^{0.685} \exp(-2.53/N^{0.0920}), & N < 16,  \\[3pt]
0.352N + 0.566 N^{0.862}, & N \geq 16.
\end{array} \right.
\label{eq:A_BCCA}
\eeq
For BPCA clusters, the averaged projected area is simply related to the radius as 
$\ovl{A}_{\rm BPCA} \approx \pi a^2$.
For more general porous aggregates, including QBCCA clusters, 
the averaged projected area is well approximated by (Equation~(47) of \citepalias{OTS09}) 
\beq
\ovl{A} = \left(\frac{1}{\ovl{A}_{\rm BCCA}(N)} + \frac{1}{\pi a^2} - \frac{1}{\pi a_{\rm BCCA}(N)^2}
\right)^{-1},
\label{eq:A}
\eeq 
where $a$ and $N$ are is the characteristic radius and monomer number of the aggregate considered,
and $a_{\rm BCCA}(N)$ and $\ovl{A}_{\rm BCCA}(N)$ are the characteristic radius and
projected area of BCCA clusters with the same monomer number $N$ (i.e., Equations~\eqref{eq:a_BCCA} 
and~\eqref{eq:A_BCCA}), respectively. 
Note that the above formula reduces to Equation~\eqref{eq:A_BCCA} 
in the BCCA limit ($a\approx a_{\rm BCCA}$) and
to $A \approx \pi a^2$ in the BPCA limit ($a \ll a_{\rm BCCA}$, $\pi a^2 \ll A_{\rm BCCA}$).

It should be noted that the above formulae 
can be only used for the average value of $A$.
This does not bother us when we compute the charge state of aggregates,
since it only depends on the {\it total} projected area $A_{\rm tot}$.
However, we cannot ignore the dispersion of $A$
when we calculate the {\it differential} drift velocity between aggregates,
especially between BCCA-like clusters.
For example, let us consider two BCCA clusters with different masses $N_1$ and $N_2(\not= N_1)$.
As Equation~\eqref{eq:A_BCCA} suggests,
the mean mass-to-area ratio $N/\ovl{A}_{\rm BCCA}$ of BCCA clusters
approaches to a constant value in the limit of large $N$.
Hence, if we ignored the area dispersion, we would have a differential drift velocity 
$\Delta u_D \propto \Delta(N/A)$ vanishing for very large $N_1$ and $N_2$ 
{\it even if} $N_1 \not= N_2$.
Clearly, this would lead to underestimation of $\Delta u_D$ and 
overestimation of the electrostatic repulsion.

To avoid this problem, we should replace $|N_1/A_1 -N_2/A_2|^2$ in ${\cal E}_D$
with $\ovl{|N_1/A_1 -N_2/A_2|^2}$, not with $|N_1/\ovl{A_1} -N_2/\ovl{A_2}|^2$,
where the overlines denote the statistical average. 
In particular, if the standard deviation of $N/A$ scales  linearly  with its mean, 
we can write $\ovl{[\Delta(N/A)]^2}$ as (see Appendix)
\beq
\ovl {\left [\Delta({N}/{A}) \right]^2} = 
\left| {N_1}/\ovl{A_1} - {N_2}/\ovl{A_2}\right|^2
+ \eps^2 \sum_{j=1,2} \left({N_j}/\ovl{A_j}\right)^2 ,
\label{eq:N/A_eps}
\eeq 
where $\eps$ is the ratio of the standard deviation to the mean of $N/A$.
In the Appendix, we evaluate $\eps$ from the numerical data on the projected area 
of sample BCCA clusters. 
We find that $\eps$ can be well approximated as $\sim 0.1$ for $N \la 10^6$.
In the following sections, we will assume $\eps = 0.1$ for all aggregates,
since the area dispersion is only important for collision between BCCA-like clusters.

\subsection{Nondimensionalization}
As seen above, our dust model is characterized by a number of model parameters.
To find a truly independent set of model parameters, 
we scale all the physical quantities involved into dimensionless ones.

We introduce the dimensionless radius and mean projected area, 
\beq
{\cal R}\equiv \frac{a}{a_0},
\eeq
\beq
{\cal A} \equiv  \frac{\ovl{A}}{\pi a_0^2}.
\eeq
Also, we  scale the mass $M$ with the the monomer number $N = M/m_0$,
where $m_0$ is the mass of monomers.
The normalized drift energy ${\cal E}_D$ and electrostatic energy ${\cal E}_E$ are
already given by Equations~\eqref{eq:ED_def} and \eqref{eq:EE_def}, respectively.
Using (${\cal R}$, ${\cal A}$, $N$) instead of ($a$, $\ovl{A}$, $M$), we have
\beq
{\cal E}_D = f_D\frac{N_1N_2}{N_1+N_2} \bigg[
\left|\frac{N_1}{{\cal A}_1}-\frac{N_2}{{\cal A}_2}\right|^2 
+ \eps^2 \sum_{j=1,2}\pfrac{N_j}{{\cal A}_j}^2
\biggr],
\label{eq:ED}
\eeq
\beq
{\cal E}_E = f_E \pfrac{\Psi}{\Psi_\infty}^2  \frac{{\cal R}_1 {\cal R}_2}{{\cal R}_1+{\cal R}_2},
\label{eq:EE}
\eeq
where the dimensionless coefficients $f_D$ and $f_E$ are defined as
\beqn
f_D & \equiv & \frac{m_0}{2 \kB T}\left(\frac{g \rho_0 a_0}{\rho_g}\sqrt{\frac{\pi m_g}{8 k_{\rm B}T}}\right)^2
\nonumber \\
&=& 1.7\times 10^{-5}\pfrac{a_0}{0.1~\micron}^{5}\pfrac{\rho_0}{1{\rm~g~cm^{-3}}}^{3}  
\nonumber \\
&&\times \pfrac{g}{10^{-3}{\rm~cm~s^{-2}}}^{2} \pfrac{\rho_g}{10^{-11}{\rm~g~cm^{-3}}}^{-2}
\pfrac{T}{100{\rm~K}}^{-2}, 
\label{eq:fD}
\eeqn
\beqn
f_E &\equiv& \frac{a_0 \Psi_\infty^2 \kB T}{e^2} 
= 0.60\Psi_\infty^2\pfrac{a_0}{0.1~\micron}\pfrac{T}{100~{\rm K}},
\label{eq:fE}
\eeqn
with the monomer material density $\rho_0 = 3m_0/4\pi a_0^3$.

We also introduce the normalized distribution function 
\beq
{\cal F}(N)dN \equiv \frac{n(M)dM}{n_0},
\eeq
where $n_0$ is the number density of monomers in the initial state.
Note that the mass conservation ensures $\int N{\cal F}(N) dN = 1$.
Using ${\cal F}$, we rewrite the ionization parameter $\Theta$ as
\beq
\Theta =  \frac{h \Psi_\infty}{{\cal A}_{\rm tot}{\cal C}_{\rm tot}},
\label{eq:Theta}
\eeq
where ${\cal A}_{\rm tot} \equiv \int {\cal A}(N){\cal F}(N) dN$ and
${\cal C}_{\rm tot} \equiv \int {\cal R}(N){\cal F}(N) dN$ are the normalized
total projected area and capacitance, and  
$h$ is a dimensionless ionization rate defined by
\beqn
h & \equiv& \frac{\zeta n_g e^2}{\pi a_0^3 n_{0}^2 \Psi_\infty \kB T}\sqrt\frac{\pi m_i}{8\kB T},
\nonumber \\
&=& 8.1\times 10^{-3} \Psi_\infty^{-1}\pfrac{a_0}{0.1~\micron}^{3}
\pfrac{\rho_0}{1{\rm~g~cm^{-3}}}^{2} 
\pfrac{\rho_d/\rho_g}{0.01}^{-2}
 \nonumber \\
&&\times \pfrac{\rho_g}{10^{-11}{\rm~g~cm^{-3}}}^{-1}\pfrac{T}{100~{\rm K}}^{-3/2}
\pfrac{\zeta}{10^{-17}~{\rm s^{-1}}}.
\label{eq:h}
\eeqn
The surface potential $\Psi$ is determined as a function of $\Theta$
by Equation~\eqref{eq:Psi}, or
\beq
\frac{1}{1+\Psi} - \frac{\exp(\Psi -\Psi_\infty) }{1+\Psi_\infty}+ 
\frac{{\cal A}_{\rm tot}{\cal C}_{\rm tot}}{h}\frac{\Psi}{\Psi_\infty} = 0,
\label{eq:Psi2}
\eeq
where we have eliminated $s_iu_i/s_eu_e$ using Equation~\eqref{eq:Psi_infty}.

From the above scaling, we find the collisional growth of charged dust aggregates
can be characterized by five dimensionless parameters ($f_D$, $f_E$, $h$, $\eps$, $\Psi_\infty$).

\section{Monodisperse Growth Model}
Before proceeding to the full simulations, we consider 
simplified situations where dust grows into {\it monodisperse} aggregates, i.e., 
where all the aggregates have the same monomer number $N$ at each moment.
This greatly helps us to understand the results of the numerical simulations shown 
in the following section. 

Within the framework of the hit-and-stick aggregation model,
the monodisperse growth is equivalent to the BCCA growth.
Thus, the assumption of the monodisperse growth 
is expressed by the following relations:
\beq
a = a_0 \pfrac{M}{m_0}^{1/D} 
\; \Longleftrightarrow \; 
{\cal R} = N^{1/D},
\label{eq:R1}
\eeq
\beq
A = A_{\rm BCCA}(N)
\; \Longleftrightarrow \; 
{\cal A} = {\cal A}(N) \equiv \frac{A_{\rm BCCA}(N)}{\pi a_0^2},
\label{eq:A1}
\eeq
\beq
n(M') = \frac{\rho_d}{M} \delta(M'-M)
\; \Longleftrightarrow \; 
{\cal F}(N') = \frac{1}{N}\delta(N-N'),
\label{eq:F1}
\eeq
where $D$ is the fractal dimension of BCCA clusters and $\delta(x)$ is the delta function.
Since $D$ is close to 2 (see Section 2.3.1), 
we simply set $D = 2$ in the following calculation.
Note that the $1/N$ factor appearing in Equation~\eqref{eq:F1} accounts for 
the mass conservation  $\int N{\cal F}(N)dN = 1$. 

Under the monodisperse approximation, the drift and electrostatic energies 
(${\cal E}_D$ and ${\cal E}_E$) can be given as a function of $N$.
Substituting Equations~\eqref{eq:R1} and \eqref{eq:A1} into Equation~\eqref{eq:ED}, 
the drift energy can be written as
\beq
{\cal E}_D =  f_D \eps^2\frac{N^3}{{\cal A}(N)^2}.
\label{eq:ED1}
\eeq
Thus, under the monodisperse approximation,  $f_D$ and $\eps$
degenerate into a single parameter $f_D\eps^2$.
Similarly, the electrostatic energy is written as 
${\cal E}_E = (f_E/2)(\Psi/\Psi_\infty)^2 N^{1/2}$, 
where $\Psi$ is given by Equation~\eqref{eq:Psi2} 
with ${\cal A}_{\rm tot} = {\cal A}(N)/N$ 
and ${\cal C}_{\rm tot} = {\cal R}/N = N^{-1/2}$.
The expression for ${\cal E}_E$ can be further simplified
using the approximate formula for $\Psi$ (Equation~\eqref{eq:Psi_approx}) 
to eliminate $\Psi/\Psi_\infty$.
The result is
\beq
{\cal E}_E = \frac{f_E}{2}
\biggl[1+ \left(h\frac{N^{3/2}}{{\cal A}(N)}\right)^{-0.8} \biggr]^{-2.5}N^{1/2}.
\label{eq:EE1}
\eeq
Note that this expression no longer involves $\Psi_\infty$. 
From Equations~\eqref{eq:ED1} and \eqref{eq:EE1},
we find that the outcome of the monodisperse growth is (approximately) determined
by three parameters $f_D\eps^2$, $f_E$, and $h$.

For later convenience, we define the ``effective kinetic energy''  ${\cal E}_K$ as
\beq
{\cal E}_K \equiv 1+ {\cal E}_D,
\label{eq:EK}
\eeq
or equivalently, $E_K \equiv {\cal E}_K \kB T = \kB T + M_\mu(\Delta u_D)^2/2$.
The first term in the right hand side of Equation~\eqref{eq:EK} 
accounts for the contribution of Brownian motion to the collisional energy ($\sim \kB T$).
We expect that the monodisperse growth is strongly suppressed
when ${\cal E}_E$ exceeds ${\cal E}_K$.

\begin{figure}
\plotone{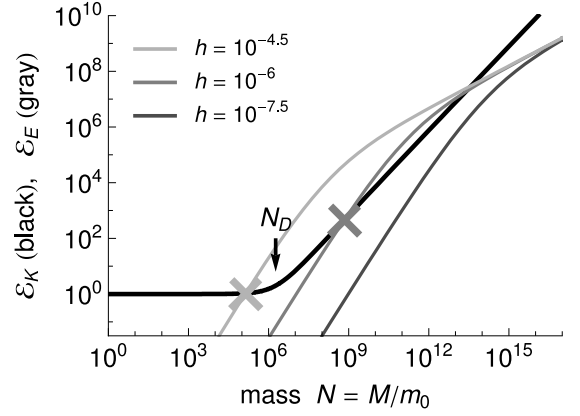}
\caption{
Examples of the effective kinetic energy ${\cal E}_K=1+{\cal E}_D$
and the electrostatic energy ${\cal E}_E$ as a function of $N$.
The black thick curve shows ${\cal E}_K$ for $f_D\eps^2 = 10^{-7}$, 
and the three gray curves show ${\cal E}_E$ for $f_E = 10$ and
$h=10^{-4.5}$, $10^{-6}$, and $10^{-7.5}$.
The black arrow shows the critical drift mass $N_D$ defined in Section~3.1,
while the gray crosses show the freezeout mass $N_F$ defined in Section~3.4
for $h =10^{-4.5}$ and $10^{-6}$. 
For $h=10^{-7.5}$, ${\cal E}_E$ is below ${\cal E}_K$ for all $N$,
so the freezeout mass is not defined.
}
 \label{fig:MDT_E}
 \end{figure}

Here, we give some examples to show how  ${\cal E}_K$ and ${\cal E}_E$ depends on
the parameters.
Figure~\ref{fig:MDT_E} shows ${\cal E}_K$ as a function of $N$ for $f_D\eps^2 =10^{-7}$.
As found from this figure, 
the kinetic energy is constant at $N \la 10^6$ due to Brownian motion (${\cal E}_K \approx 1$),
 and increases with mass at  $N \ga 10^6$
 due to the differential drift (${\cal E}_K \approx {\cal E}_D \propto N^3/{\cal A}^2$) .
The qualitative behavior is the same for every $f_D\eps^2$.
The value of $f_D\eps^2$ only determines the mass at which 
the differential drift starts to dominate over Brownian motion in the kinetic energy.
In figure~\ref{fig:MDT_E}, we also plot ${\cal E}_E$ for $f_E = 10$
with varying the value of $h (=10^{-4.5}$, $10^{-6}$, $10^{-7.5}$).
For all the cases, ${\cal E}_E$ quickly increases with $N$ and finally 
becomes proportional to ${\cal R} = N^{1/2}$. 
This reflects the transition of the plasma state from 
the IDP ($\Psi \approx \Theta \propto N^{3/2}/{\cal A}$) to the IEP ($\Psi \approx \Psi_\infty$).
In the IEP limit, ${\cal E}_E$ depends on $f_E$ but is independent of $h$.
An important difference among the three examples is the timing of the plasma transition:
for smaller $h$, ${\cal E}_E$ approaches the IEP limit at larger $N$.
This difference makes the ratio between ${\cal E}_E$ and ${\cal E}_K$ 
qualitatively different among the three cases.
For $h = 10^{-4.5}$, ${\cal E}_E$ exceeds ${\cal E}_K$ when 
the relative motion is dominated  by Brownian motion.
For $h = 10^{-6}$, by contrast, ${\cal E}_E$ exceeds ${\cal E}_K$ when 
the relative motion is dominated by the differential drift.
For $h = 10^{-7.5}$, ${\cal E}_E$ does not exceed ${\cal E}_K$ for arbitrary $N$.
As we see in Section 4, this difference is a key to understand the collisional growth of 
dust aggregates with size distribution.

To quantify these differences for general cases,  we introduce the following quantities:
\begin{itemize}
\item {\it The drift mass} $N_D$.
This is defined as the mass at which the relative motion
starts to be  dominated by the differential drift.
\item {\it The plasma transition mass} $N_P$.
This is defined as the mass at which the plasma state shifts from the IDP to the IEP.  
\item {\it The maximum energy ratio} $({\cal E}_E/{\cal E}_K)_{\rm max}$.
This is the maximum value of the ratio ${{\cal E}_E/{\cal E}_K}$ in the monodisperse growth.
If $({\cal E}_E/{\cal E}_K)_{\rm max} >1$,
the electrostatic energy ${\cal E}_E$ exceeds the kinetic energy ${\cal E}_K$ at a certain mass.
\item {\it The freezeout mass} $N_{F}$.
This is  the mass at which ${\cal E}_E$ starts to exceed ${\cal E}_K$.
Note that the freezeout mass is only defined when $({\cal E}_E/{\cal E}_K)_{\rm max} >1$. 
\end{itemize}
In the following subsection, we describe how these quantities are related
 to the parameters ($f_D\eps^2$, $f_E$, $h$). 
 
\subsection{$N_D$: the  Drift Mass}
The first and second terms in the right hand side of Equation~\eqref{eq:EK}
represents Brownian motion and the differential drift.
Since the second term monotonically increases with $N$, there exists a critical mass
at which the dominant relative motion changes from the Brownian motion to the differential drift.
We define $N_D$ as the critical mass satisfying ${\cal E}_D(N_D) = 1$.
Using Equation~\eqref{eq:ED1}, the equation for $N_D$ is written as
\beq
\frac{{\cal A}(N_D)^2}{N_D^3} = f_D \eps^2.
\label{eq:ND}
\eeq
This equation implicitly determines $N_D$ as a function of $f_D\eps^2$.
For example, $N_D \approx 3\times 10^6$ when $f_D\eps^2 = 10^{-7}$ 
(see Figure~\ref{fig:MDT_E}).

\begin{figure}
\plotone{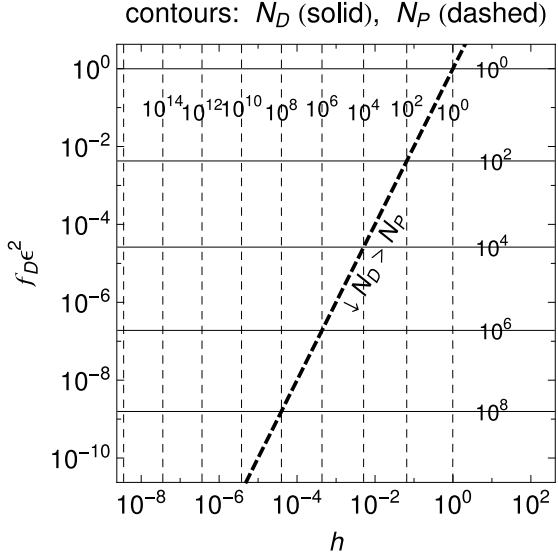}
\caption{
Contour plot of the drift mass $N_D$ (Equation~\eqref{eq:ND}; solid lines) 
and the plasma transition mass $N_P$ (Equation~\eqref{eq:NP}; dashed lines)
as a function of $h$ (x-axis) and $f_D\eps^2$ (y-axis).
}
\label{fig:NDNP}
\end{figure}
\begin{figure}
\epsscale{0.8}
\plotone{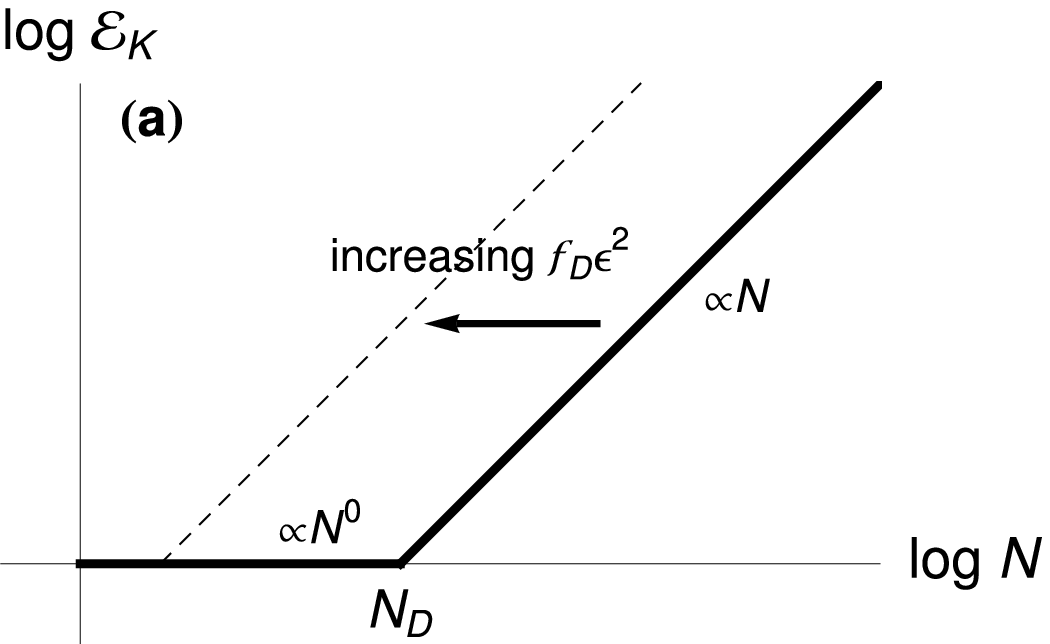}
\vspace{3mm}
\plotone{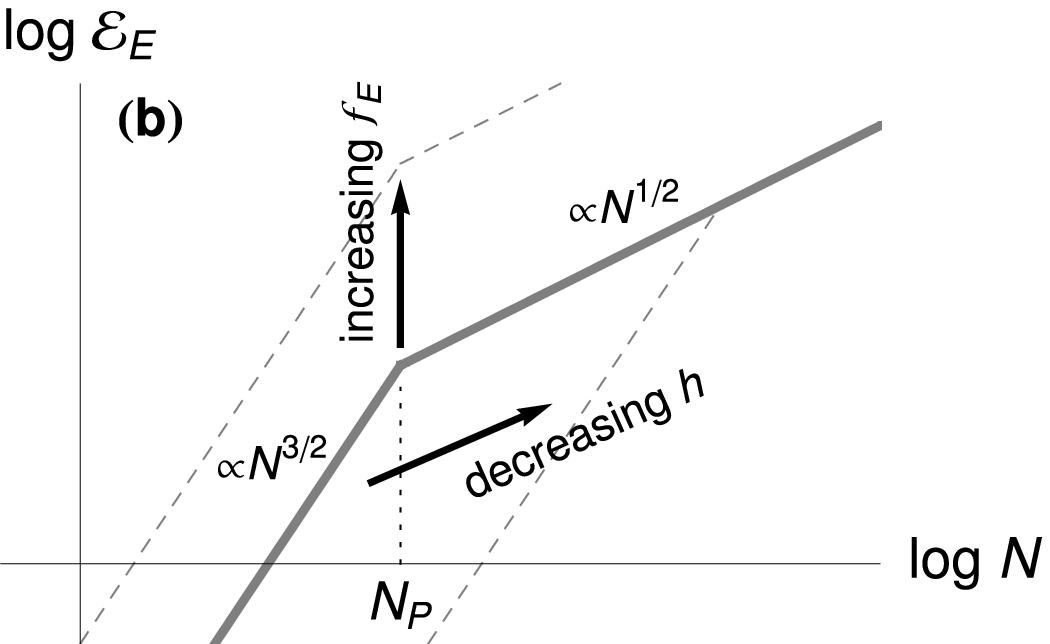}
\caption{
Schematic diagrams describing the mass dependence 
of the effective kinetic energy ${\cal E}_K$ (a) 
and the electrostatic energy ${\cal E}_E$ (b).
Here, $N_D$ and $N_P$ are the  drift mass and plasma transition mass defined by
Equations~\eqref{eq:ND} and \eqref{eq:NP}, respectively.
The dashed lines with arrows indicate how ${\cal E}_K$ and ${\cal E}_E$ depends 
on the parameters $f_D \eps^2$, $f_E$, and $h$.
}
\label{fig:EKEE}
\end{figure}

Figure \ref{fig:NDNP} shows the solution to Equation~\eqref{eq:ND}.
When $f_D\eps^2 \ll 1$,  $N_D$ is well approximated as
\beq
N_D \approx \frac{1}{b^2 f_D\eps^2},
\label{eq:ND_approx}
\eeq
where $b=1/0.352=2.84$ is the mass-to-area ratio $N/{\cal A}(N)$ in the limit of $N\to\infty$.
Using Equation~\eqref{eq:ND_approx}, ${\cal E}_K$ is simply rewritten as
\beq
{\cal E}_K \approx 1+ \frac{N}{N_D},
\label{eq:EK_approx}
\eeq
which asymptotically behaves as ${\cal E}_K \approx 1$ for $N\ll N_D$ and 
${\cal E}_K \approx N/N_D$ for $N\gg N_D$. 
The asymptotic form of ${\cal E}_K$ is schematically illustrated in Figure~\ref{fig:EKEE}(a).

\subsection{$N_P$: the Plasma Transition Mass}
Another important quantity is the critical mass at which the 
plasma state changes from IDP to IEP.
We define the critical mass $N_P$ such that $\Theta(N_P) = \Psi_\infty$
 (see Equation~\eqref{eq:Psi_asympt}).
Using Equation~\eqref{eq:Theta}, this condition can be written as
\beq
\frac{{\cal A}(N_P)}{N_P^{3/2}} = h.
\label{eq:NP}
\eeq
Note that $N_P$ depends on $h$ only.

Figure~\ref{fig:NDNP} shows the solution to Equation~\eqref{eq:NP}
as a function of $h$.
If $h \ll 1$,  $N_P$ is well approximated as
\beq
N_P \approx \frac{1}{b^2 h^2}.
\label{eq:NP_approx}
\eeq
In this case, ${\cal E}_E$ can be approximately written as
\beq
{\cal E}_E \approx \frac{f_E}{2} \biggl[ 1+\pfrac{N}{N_P}^{-0.4} \biggr]^{-2.5} N^{1/2}
\label{eq:EE_approx}
\eeq
which asymptotically behaves as  
${\cal E}_E \approx (f_E/2)N^{3/2}/N_P$ for $N \ll N_P$
and as ${\cal E}_E \approx (f_E/2)N^{1/2}$ for $N \gg N_P$.
The asymptotic form of ${\cal E}_E$ is illustrated in Figure~\ref{fig:EKEE}(b).

\subsection{$({\cal E}_E/{\cal E}_K)_{\rm max}$: the Maximum Energy Ratio}
The maximum energy ratio $({\cal E}_E/{\cal E}_K)_{\rm max}$
determines whether the electrostatic energy exceeds the kinetic energy
during the monodisperse growth.
Since ${\cal E}_E$ scales linearly with $f_E$, 
the quantity $f_E^{-1}({\cal E}_E/{\cal E}_K)_{\rm max}$
depends only on $f_D\eps^2$ and $h$.

\begin{figure}
\plotone{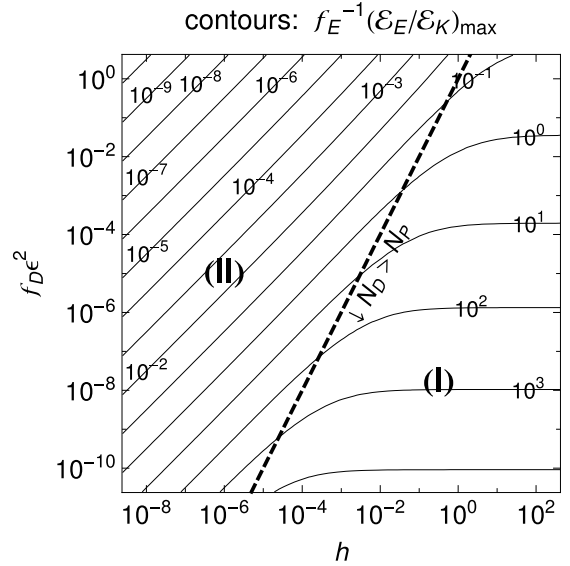}
\caption{
Contour plot of the maximum energy ratio $({\cal E}_E/{\cal E}_K)_{\rm max}$ 
divided by $f_E$ as a function of $h$ (x-axis) and $f_D \eps^2$ (y-axis).
The dashed line represents $N_D = N_P$ (see also Figure~\ref{fig:NDNP}).
The two parameter regions (I) and (II) are characterized by $N_D \gg N_P$ and $N_D \ll N_P$,
respectively (see also Figure~\ref{fig:EKEE_Ermax}). 
}
\label{fig:Ermax}
\end{figure}
\begin{figure}
\epsscale{0.8}
\plotone{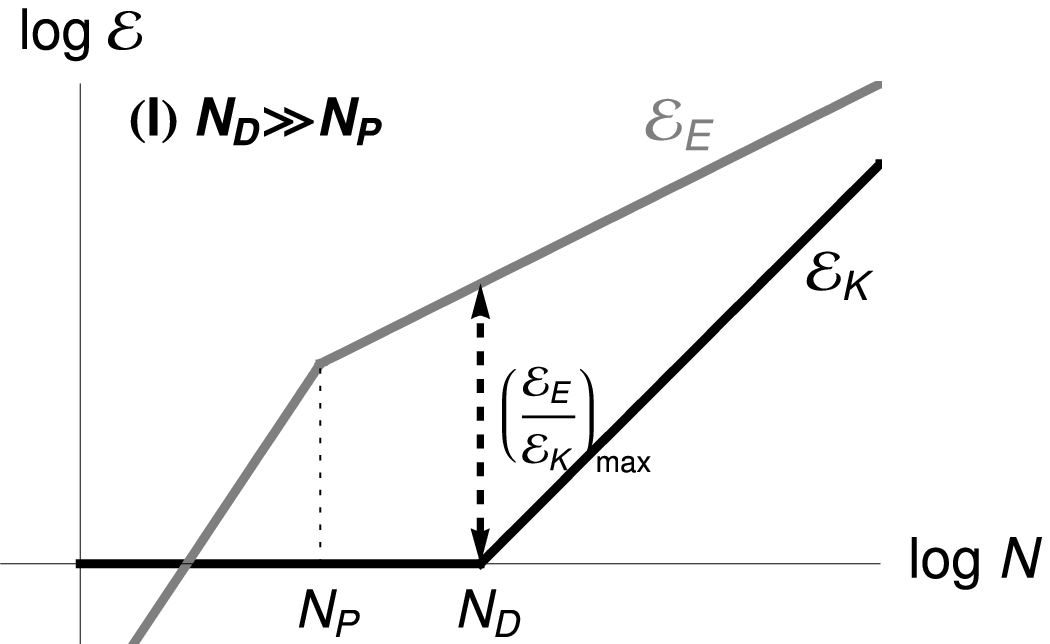}
\vspace{3mm}
\plotone{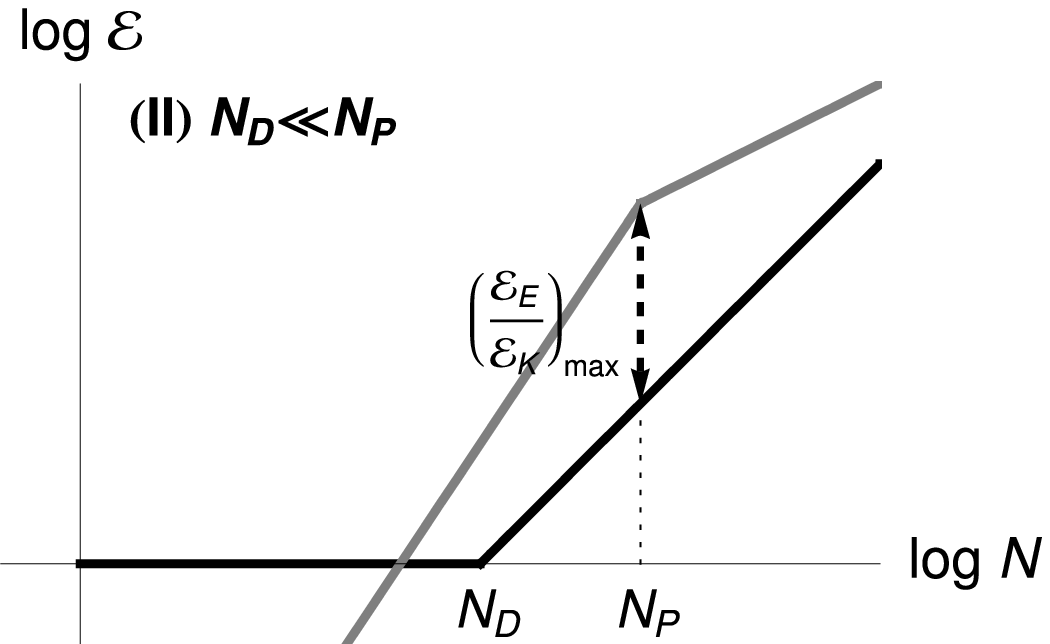}
\caption{
Schematic diagrams describing the dependence of the maximum energy ratio
$({\cal E}_E/{\cal E}_K)_{\rm max}$ on $N_D$ and $N_P$ in regions (I) and (II)
shown in Figure~\ref{fig:Ermax}.
The black and gray lines shows the asymptotic behavior of 
 ${\cal E}_K$ and ${\cal E}_E$ (Equations~\eqref{eq:EK_approx} and \eqref{eq:EE_approx})
  as a function of $N$, respectively.
When $N_D \gg N_P$, or equivalently $f_D\eps^2 \ll h^2$ (region (I); upper panel), 
the energy ratio maximizes at $N \approx N_D$.
In the opposite limit (region (II); lower panel), 
${\cal E}_E/{\cal E}_K$ maximizes at $N \approx N_P$. 
}
\label{fig:EKEE_Ermax}
\end{figure}

Figure~\ref{fig:Ermax} plots $f_E^{-1}({\cal E}_E/{\cal E}_K)_{\rm max}$ 
 as a function of $f_D\eps^2$ and $h$. 
It is seen that the maximum energy ratio behaves differently across the line $N_D = N_P$.
This can be easily understood from Figure~\ref{fig:EKEE_Ermax},
which schematically illustrates the mass dependence of ${\cal E}_K$ and ${\cal E}_E$ 
(Equations~\eqref{eq:EK_approx} and \eqref{eq:EE_approx}).
If $N_D \gg N_P$ ($f_D\eps^2 \ll h^2$), the energy ratio ${\cal E}_E/{\cal E}_K$ 
reaches the maximum at $N \approx N_D$.
Since ${\cal E}_D(N_D) = 2$ and ${\cal E}_E(N_D) \approx f_E N_D^{1/2}/2$,
we obtain $({\cal E}_E/{\cal E}_K)_{\rm max} \approx f_E N_D^{1/2}/4 \approx f_E/4b f_D^{1/2}\eps$
independently of $h$.
If $N_D \ll N_P$ ($f_D\eps^2 \gg h^2$), by contrast, ${\cal E}_E/{\cal E}_K$
reaches the maximum at $N \approx N_P$.
Using ${\cal E}_D(N_P) = N_P/N_D$ and ${\cal E}_E(N_P) \approx f_E/2^{3.5}$,
we have $({\cal E}_E/{\cal E}_K)_{\rm max} \approx f_E N_D/2^{3.5}N_P
\approx f_Eh/2^{3.5}bf_D\eps^2$, which depends on both $f_D\eps^2$ and $h$.

\subsection{$N_F$: the Freezeout Mass} 
When $({\cal E}_E/{\cal E}_K)_{\rm max}>1$, there exists a critical mass $N_F$
at which the electrostatic energy ${\cal E}_E$ takes over the kinetic energy ${\cal E}_K$.
As we will see in Section 3.5, the monodisperse growth is strongly suppressed at  $N \ga N_F$.
For this reason, we refer to $N_F$ as the ``freezeout mass.'' 
\begin{figure}
\plotone{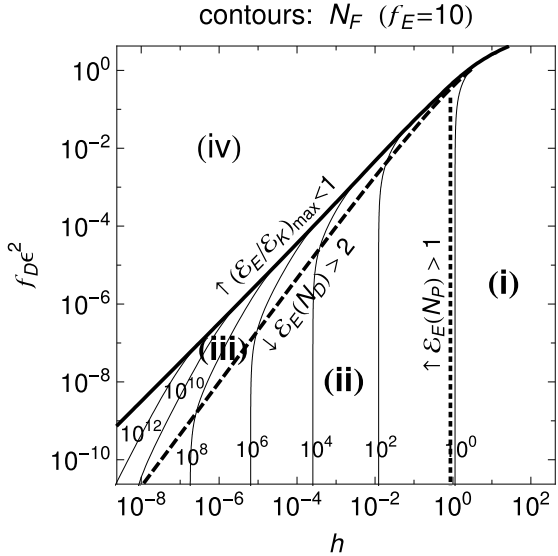}
\caption{
Contour plot of the freezeout mass $N_F$ (thin solid curves) 
for $f_E = 10$ as a function of $h$  (x-axis) and $f_D \eps^2$ (y-axis).
The dashed and dotted curves show ${\cal E}_E(N_D) = 2$ and ${\cal E}_E(N_P) = 1$,
respectively. 
The regions (i), (ii), and (iii) are characterized by the values of 
${\cal E}_E(N_D)$ and ${\cal E}_E(N_P)$ (see also Figure~\ref{fig:EKEE_NF}).
Above the thick solid curve (region (iv)), 
the maximum energy ratio $({\cal E}_E/{\cal E}_K)_{\rm max}$ is less than unity,
so the freezeout mass is not defined.
}
\label{fig:NF}
\end{figure}
\begin{figure}
\epsscale{0.8}
\plotone{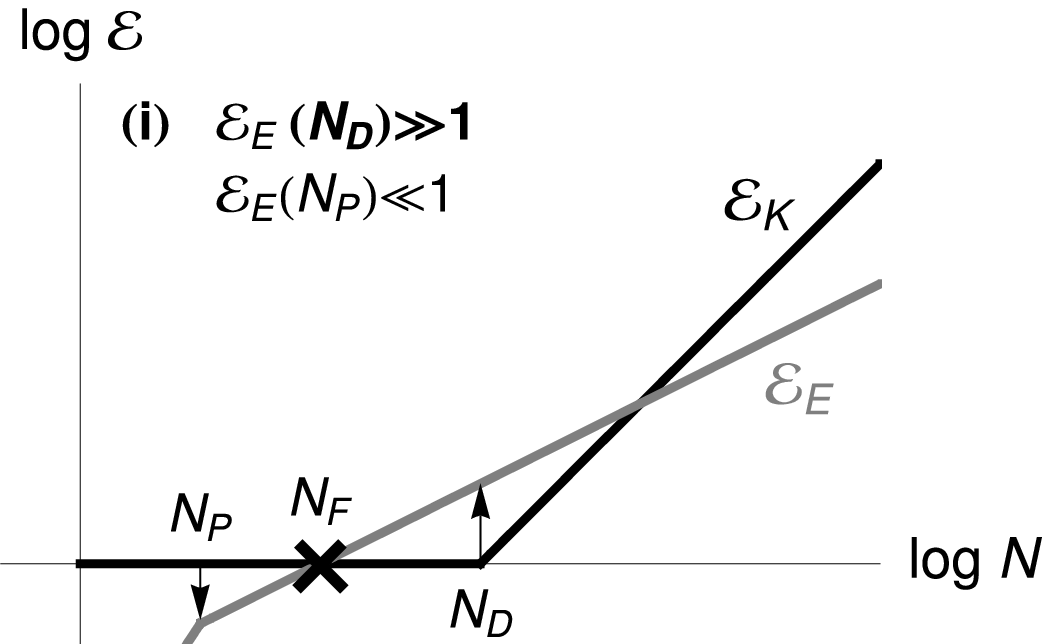}
\vspace{3mm}
\plotone{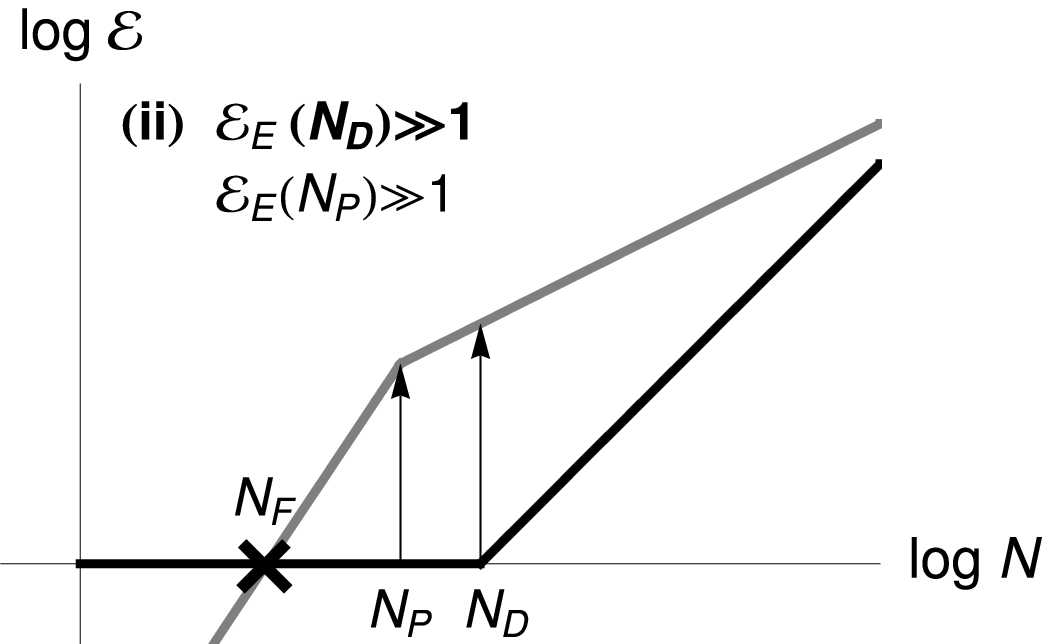}
\vspace{3mm}
\plotone{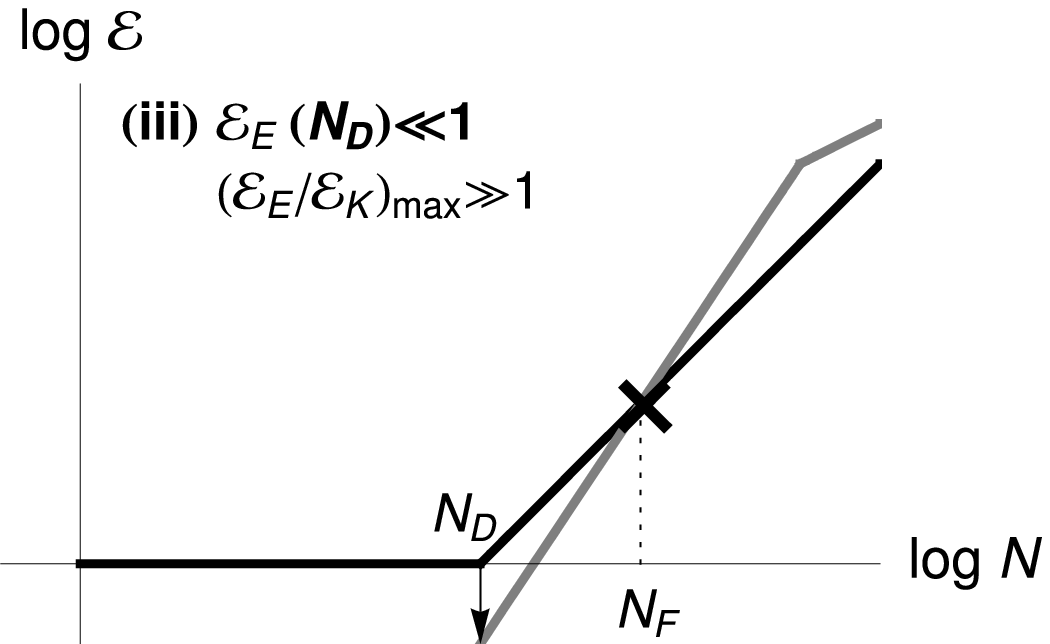}
\epsscale{1.0}
\caption{
Schematic diagrams describing the location of the freezeout mass $N_F$ in the mass space  
for three parameter regions (i), (ii) and (iii) shown in Figure~\ref{fig:NF}.
The black and gray lines shows the asymptotic behavior of 
 ${\cal E}_K$ and ${\cal E}_E$ (Equations~\eqref{eq:EK_approx} and \eqref{eq:EE_approx})
  as a function of $N$, respectively.
If  ${\cal E}_E(N_D) \gg 1$ (cases (i) and (ii); top and middle panels),
${\cal E}_E$ exceeds ${\cal E}_K$ in the Brownian motion regime (i.e., $N_F \ll N_D$).
If  ${\cal E}_E(N_D) \ll 1$ but still $({\cal E}_E/{\cal E}_K)_{\rm max} \gg 1$ (case (iii); bottom panel),
${\cal E}_E$ exceeds ${\cal E}_K$ in the differential drift regime (i.e., $N_F \gg N_D$).
}
\label{fig:EKEE_NF}
\end{figure}
The freezeout mass can be calculated from the condition ${\cal E}_K(N_F) = {\cal E}_E(N_F)$ 
once the three parameters $f_D\eps^2$, $f_E$, and $h$ are specified.

In Figure~\ref{fig:NF}, we plot $N_F$ as a function of $f_D\eps^2$ and $h$ for $f_E = 10$.
We see that $N_F$ depends on these parameters differently depending on
the values of ${\cal E}_E(N_P)$ and ${\cal E}_E(N_D)$.
To understand this, in Figure~\ref{fig:EKEE_NF}, 
we schematically show ${\cal E}_K$ and ${\cal E}_E$ as a function of $N$ for the three cases.
If ${\cal E}_E(N_D) \gg 1$,  ${\cal E}_E$ starts to exceed ${\cal E}_K$ when  
the relative velocity is dominated by Brownian motion (i.e., $ N_F\ll N_D$).
In this case, the condition determining $N_F$ is given by $ {\cal E}_E(N_F) \approx 1$,
which implies $N_F \approx (2/f_E)^2$ for ${\cal E}(N_P) \ll 1$
and  $N_F \approx (2N_P/f_E)^{2/3} \approx (2/b^2 f_E h^2)^{2/3}$ for ${\cal E}(N_P) \gg 1$.
If ${\cal E}_E(N_D) \ll 1$ but still $({\cal E}_E/{\cal E}_K)_{\rm max} \gg 1$,  
${\cal E}_E$ exceeds ${\cal E}_K$ after the relative velocity is 
dominated by the differential drift (i.e., $ N_F\gg N_D$).
In this case, the condition for $N_F$ is given by $(f_E/2)N_F^{3/2}/N_P \approx N_F/N_D$,
hence $N_F$ is given by $N_F \approx (2N_P/f_EN_D)^{2} \approx (2 f_D\eps^2/f_E h^2)^{2}$.

\subsection{The Outcomes of Monodisperse Growth}
As mentioned above, the monodisperse growth is expected to slow down
at the freezeout mass $N \approx N_F$ when $({\cal E}_E/{\cal E}_K)_{\rm max}>1$.
Here, we demonstrate this by numerically calculating the mass evolution.

Under the monodisperse approximation, the evolution of aggregate mass $N$ is given by
\beq
\frac{dM}{dt} = \rho_d K 
\; \Longleftrightarrow \;
\frac{dN}{d{\cal T}} = {\cal K}
\label{eq:N_evol}
\eeq
where ${\cal T} = n_0\pi a_0^2 t\sqrt{8 \kB T/\pi m_0}$ 
and ${\cal K} = K/(\pi a_0^2 \sqrt{8 \kB T/\pi m_0})$ are the 
dimensionless time and collisional rate coefficient.
We numerically solve Equation~\eqref{eq:N_evol} with initial condition $N({\cal T}=0) = 1$.

As in the beginning of this section, we consider three cases of $h = 10^{-4.5}$, $10^{-6}$, 
and $10^{-7.5}$ with fixed $f_D\eps^2 = 10^{-7}$ and $f_E = 10$.
Listed in Table~\ref{table1} are the critical masses ($N_D$, $N_P$, $N_F$) 
and the maximum energy ratio $({\cal E}_E/{\cal E}_K)_{\rm max}$ for these cases.
We also consider the uncharged case with the same value of  $f_D\eps^2$.
\begin{deluxetable}{lllll}
\tablecaption{Critical Masses and the Maximum Energy Ratio 
for $(f_D\eps^2,f_E)=(10^{-7},10)$}
\tablecolumns{5}
\tablehead{
\colhead{$h$} & \colhead{$N_D$} &
\colhead{$N_P$} & \colhead{$({\cal E}_E/{\cal E}_K)_{\rm max}$} &
\colhead{$N_F$} 
}
\startdata
$10^{-4.5}$ & $10^{6.3}$ &$10^{7.2}$  & $10^{2.0}$ & $10^{5.1}$ \\
$10^{-6}$ & $10^{6.3}$ &$10^{10.1}$  & $10^{0.5}$ & $10^{8.8}$ \\
$10^{-7.5}$ & $10^{6.3}$ &$10^{13.1}$  & $10^{-1.0}$ & \nodata
\enddata
\label{table1}
\end{deluxetable}

\subsubsection{Without Charging}
\begin{figure}
\plotone{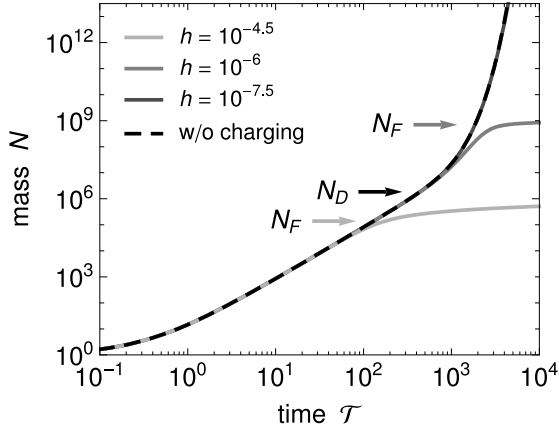}
\caption{
Mass evolution in the monodisperse model calculated from  Equation~\eqref{eq:N_evol}
for $f_D\eps^2 = 10^{-7}$ and $f_E = 10$ with various values of $h$.
The black arrow indicate the drift mass $N_D$, 
while the lower and upper arrows show the freezeout mass $N_F$
 for $h=10^{-4.5}$ and $10^{-6}$, respectively.
The evolution for the uncharged case (i.e., $h = 0$) is shown by the dashed curve.  
}
 \label{fig:MDT_tN}
 \end{figure}

In Figure~\ref{fig:MDT_tN}, the mass evolution for the uncharged case is shown 
by the dashed curve. 
The black arrow in the figure indicates the critical drift mass $N_D = 10^{6.3}$. 
We find that the mass grows as ${\cal T}^2$ until reaching $N_D$, and then 
grows exponentially with ${\cal T}$.
This evolutionary trend can be directly proven from Equation~\eqref{eq:N_evol}.
Without charging, the collision kernel ${\cal K}$ is just the product 
of the geometrical cross section $\propto {\cal R}^2 = N$ and the relative velocity $\Delta u$. 
When $N \ll N_D$, the relative velocity is dominated by Brownian motion 
(i.e., $\Delta u \propto N^{-1/2}$), and we have 
${\cal K} \propto  {\cal R}^2N^{-1/2}\propto N^{1/2}$.
Inserting this into Equation~\eqref{eq:N_evol}, we have $N \propto {\cal T}^2$.
When $N \gg N_D$, by contrast, the relative velocity is dominated by the differential drift 
($\Delta u \propto N/{\cal A}$), and hence ${\cal K} \propto N{\cal R}^2/{\cal A}$.
Since the projected area ${\cal A}$ roughly scales with ${\cal R}^2$,  
we have ${\cal K} \propto N$.
Hence, from Equation~\eqref{eq:N_evol}, we find $N \propto \exp(\Omega {\cal T})$,
where $\Omega$ is a constant growth rate.

\subsubsection{With Charging}
The mass evolution for the charged cases is plotted in Figure~\ref{fig:MDT_tN} by gray curves.
The gray arrows in the figure indicate the freezeout mass $N_F$ for $h=10^{-4.5}$ and $10^{-6}$ . 
As expected, we observe significant slowdown in the growth at $N \approx N_F$ for the two cases.
At ${\cal T} = 10^4$, the aggregate mass is $N\approx 10^{5.7}$ for $h=10^{-4.5}$ and 
$N\approx 10^{8.9}$ for $h = 10^{-6}$, which is consistent with the predicted freezeout mass
(see Table~\ref{table1}). 
We have computed the mass evolution for the two cases until ${\cal T}=10^6$, 
but the final masses $10^{5.9}$ and $10^{9.0}$ are not very different from the values at ${\cal T}=10^4$.

For $h = 10^{-7.5}$, by contrast, the evolution curve of $N$ is indistinguishable from 
that for the uncharged case, meaning that the electrostatic repulsion hardly affects the aggregate growth.

To summarize, we have confirmed that dust can continue the monodisperse growth only if 
\beq
\pfrac{{\cal E}_E}{{\cal E}_K}_{\rm max} \la 1.
\label{eq:growthcond_MDT}
\eeq

\section{Numerical Simulations Including Size Distribution}
As shown in the previous section,
dust aggregates could not grow beyond the freezeout mass $N_F$ 
if the condition \eqref{eq:growthcond_MDT} is not satisfied
{\it and} if the size distribution were limited to monodisperse ones.  
In this section, we study how the outcome of dust growth changes 
when we allow the size distribution to freely evolve. 
 
To compute the evolution of size distribution,
we employ the ``extended'' Smoluchowski method developed in \citetalias{OTS09}.
This method treats the number density $n(M)$ and the {\it mean volume} $\ovl{V}(M)$ of
aggregates with mass $M$ as time-dependent quantities, 
and calculates their temporal evolution simultaneously.
This method allows us to follow the porosity evolution consistently with collisional growth, 
which cannot be done with the conventional Smoluchowski method \citep[e.g.,][]{NNH81,THI05,DD05}. 

In the extended Smoluchowski method, the temporal evolution of $n(M)$ and $\ovl{V}(M)$ 
is given by two equations,
\beqn
\frac{\pd n(M)}{\pd t} 
&=& \frac{1}{2}\int_0^M dM' \; \overline{K}(M';M-M')n(M')n(M-M') \nonumber \\
&& - n(M)\int_0^\infty dM'\; \overline{K}(M;M')n(M'),
\label{eq:smol_n}
\eeqn
\beqn
\frac{\pd[ \ovl{V}(M) n(M)]}{\pd t} 
&=& \frac{1}{2}\int_0^M dM' \; 
\overline{V}_{1+2}(M';M-M')\overline{K}(M';M-M')\nonumber \\
&&\qquad \times n(M')n(M-M') \nonumber \\
&& - \ovl{V}(M)n(M)\int_0^\infty dM' \; \overline{K}(M;M')n(M'), 
\label{eq:smol_Vn}
\eeqn
where $\overline{K}(M_1;M_2)$ and $\ovl{V}_{1+2}(M_1;M_2)$ 
are the collisional rate coefficient $K$ (Equation~\eqref{eq:K}) 
and the aggregate volume $V_{1+2}$ after a collision (Equation~\eqref{eq:V_1+2}) 
evaluated for $V_1 = \ovl{V}(M_1)$ and $V_2 = \ovl{V}(M_2)$. 
In this study, we determine $V_{1+2}$ using the formula for hit-and-stick collisions
(Equation~\eqref{eq:chi}). 

We numerically solve Equations~\eqref{eq:smol_n} and \eqref{eq:smol_Vn} 
using the fixed bin scheme described in \citetalias{OTS09}.
This scheme divides the low-mass region $m_0 \leq M \leq {\cal N}_{bd}m_0$ into linearly spaced bins 
with representative masses $M_k = km_0~(k=1,2,\dots,{\cal N}_{bd})$ and the high-mass region
$M > {\cal N}_{bd}m_0$ into logarithmically spaced bins with 
$M_k = 10^{1/{\cal N}_{bd}}M_{k-1}~(k={\cal N}_{bd}+1,\dots)$.
The number ${\cal N}_{bd}$ controls the resolution in the mass coordinate.
In this study, we set ${\cal N}_{bd} = 80$ (meaning $M_{k+1}/M_k = 1.03$
for the high-mass range).
The temporal evolution is computed using the explicit, forth-order Runge--Kutta method.
The time increment $\Delta t$ for each time step is continuously adjusted so that
the fractional decrease in the number density during $\Delta t$ 
does not exceed $\delta_t$ for all bins, where $\delta_t$ is a constant parameter.
We take $\delta_t = 0.02$ in the following calculations.

\subsection{Without Charging}
 \begin{figure}
\plotone{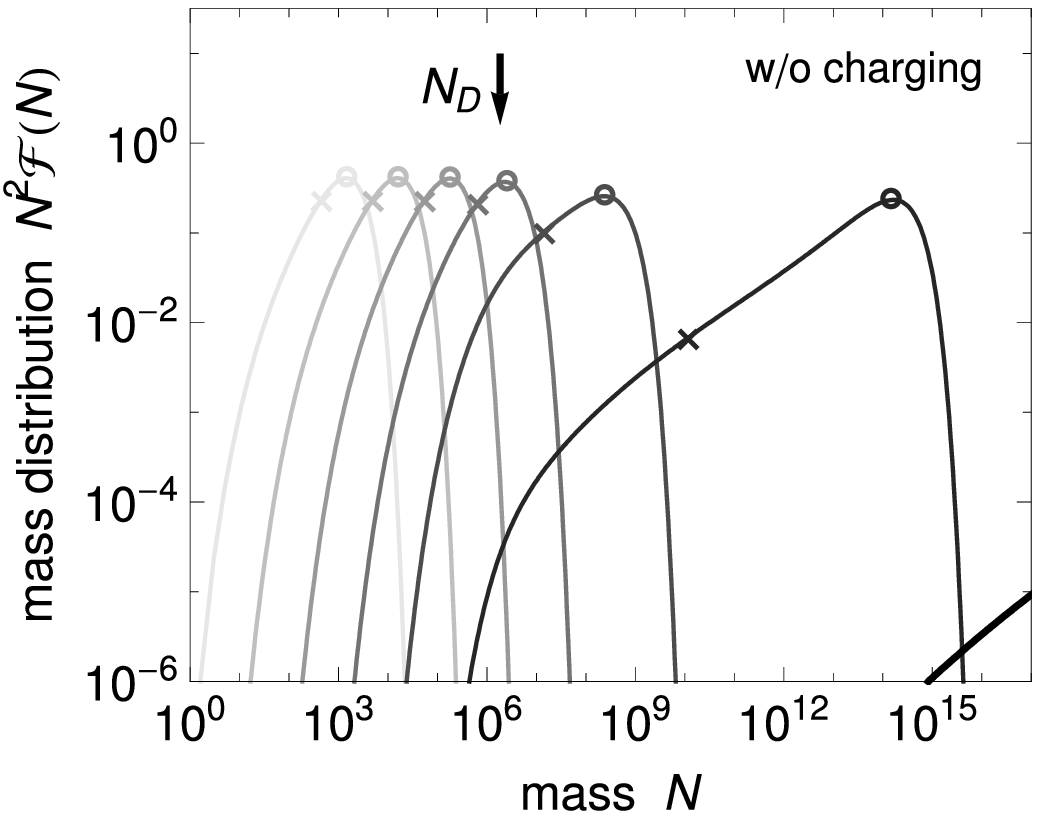}
\vspace{2mm}
\plotone{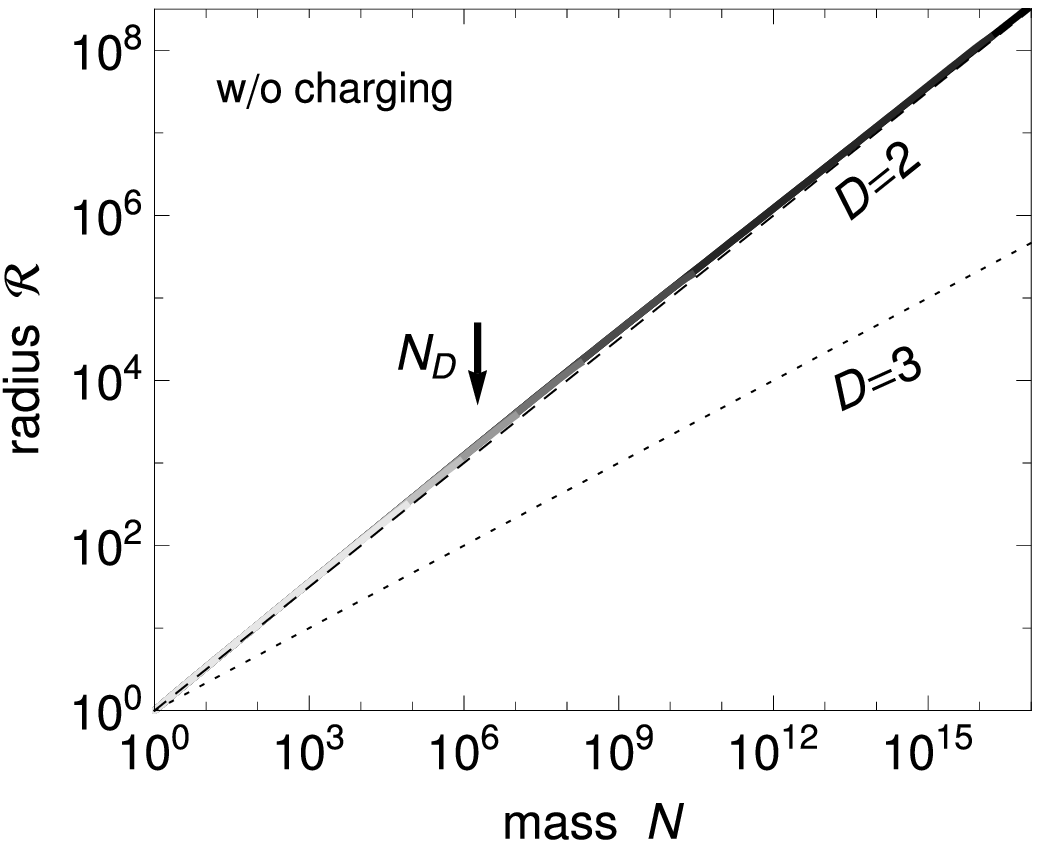}
 \caption{
Evolution of the mass distribution function ${\cal F}(N)$ (upper panel) and 
the mass--radius relation ${\cal R}(N)$ (lower panel) 
for the uncharged case of $(f_D\eps^2, \eps) = (10^{-7},10^{-1})$.
The gray curves show the snapshots of $N^2{\cal F}(N)$ and ${\cal R}(N)$
 at various times, ${\cal T} = 10^1$, $10^{1.5}$, $10^2, \dots,10^{4}$ (from left to right).  
Note that the curves for ${\cal R}(N)$ overlap each other.
The arrows indicate the critical mass $N_D$ calculated from the monodisperse 
theory (Equation~\eqref{eq:ND}).
The crosses and open circles in the upper panel indicate the averaged mass $\bracket{N}$ 
(Equation~\eqref{eq:Navr}) and the {\it weighted} averaged mass $\bracket{N}_m$ 
(Equation~\eqref{eq:Navrm}), respectively.
In the lower panel, the mass--radius relations for the fractal dimensions of $D= 2$ and 3 are
shown by the dashed and dotted curves, respectively.
}
 \label{fig:evol0}
 \end{figure}
Figure~\ref{fig:evol0} shows the solution to Equations~\eqref{eq:smol_n} and \eqref{eq:smol_Vn}
for the uncharged case of $(f_D\eps^2, \eps)=(10^{-7},10^{-1})$. 
The upper panel displays the mass distribution function ${\cal F}(N)$ at various times ${\cal T}$.
Note that the vertical axis of this panel is chosen to be $N^2{\cal F}(N)$, 
which is proportional to the mass density of aggregates belonging to each logarithmic mass bin.

To characterize the evolution of the mass distribution, we introduce
 the average mass $\bracket{N}$ and the {\it mass-weighted} average mass $\bracket{N}_m$ defined by
\beq
\bracket{N} \equiv \frac{ \int_0^\infty N{\cal F}(N)dN}{\int_0^\infty {\cal F}(N)dN} 
=  \frac{1}{\int_0^\infty {\cal F}(N)dN}, 
\label{eq:Navr}
\eeq
\beq
\bracket{N}_m \equiv \frac{ \int_0^\infty N^2{\cal F}(N)dN}{\int_0^\infty N{\cal F}(N)dN} 
= \int_0^\infty N^2{\cal F}(N)dN,
\label{eq:Navrm}
\eeq
where we have used the mass conservation $\int_0^\infty N{\cal F}(N) dN = 1$.
Note that $\bracket{N}$ is inversely proportional to the total number density of aggregates,
$\int_0^\infty {\cal F}(N) dN$. 
Roughly speaking,  $\bracket{N}$ represents the mass scale dominating the number of aggregates 
in the system, while $\bracket{N}_m$ represents the mass scale dominating the mass of the system.
Also note that $\bracket{N}_m$ can be written as $\bracket{N^2}/\bracket{N}$,
and the dispersion $\bracket{\delta N^2} \equiv \bracket{N^2} - \bracket{N}^2$
 of the mass distribution is written as 
 $\bracket{\delta N^2} = \bracket{N}^2(\bracket{N}_m/\bracket{N} -1)$.
Hence, the ratio $\bracket{N}_m / \bracket{N}$ measures  
how the mass distribution deviates from the monodisperse distribution.
In the upper panel of Figure~\ref{fig:evol0}, we indicate $\bracket{N}$ and $\bracket{N}_m$
at each time with crosses ($\times$) and circles ($\circ$), respectively.

The evolution of the mass distribution can be divided into two stages.
During $\bracket{N}_m \la N_D$, 
the mass distribution evolves with small dispersion ($\bracket{N} \approx \bracket{N}_m$).
The average masses $\bracket{N}$ and $\bracket{N}_m$ grow approximately as ${\cal T}^2$,
which is consistent with the prediction of the monodisperse theory (see Section~3.5.1).
These imply that the monodisperse approximation is good
when Brownian motion dominates the relative motion of aggregates.

However, the monodisperse approximation becomes less good once $\bracket{N}_m$ exceeds $N_D$.
In this stage, we observe a power-law tail extending from $N \approx \bracket{N}_m$ 
down to $N \approx N_D$.
In fact, we see that the growth rate of  $\bracket{N}_m$ 
(i.e., $d\ln\bracket{N}_m/d{\cal T}$) is approximately twice as high as that of $\bracket{N}$. 
This means that  the relative width of the distribution 
($\bracket{\delta N^2}^{1/2}/\bracket{N} = \sqrt{\bracket{N}_m/\bracket{N} - 1}
\approx \sqrt{\bracket{N}_m/\bracket{N}}$)
increases exponentially with time\footnote{
As pointed out by the referee, this is a general consequence of the kernel ${\cal K}$ 
scaling {\it linearly} with the masses of colliding aggregates 
(this is the case for our kernel at $N \gg N_D$, see Section 3.5.1).
In fact, the growth rate of $\bracket{N}_m$ is known to be 
{\it exactly} twice as high as that of $\bracket{N}$ 
when the kernel is of the form ${\cal K}(N_1;N_2) \propto N_1 + N_2$ 
\citep[see, e.g., Figure~1 of][]{OS08}.
}.
As we will see in the following subsection, the broadening of the mass distribution 
plays a key role when dust charging is present.

The lower panel of Figure~\ref{fig:evol0} shows the temporal evolution of
the mass-radius relation ${\cal R}(N)$. We see that ${\cal R}(N)$ approximately obeys 
a fractal relation ${\cal R} \approx N^{1/D}$, where the fractal dimension is $D \approx 2$
 independently of the time (see the dashed line in the panel which shows 
 the exact relation ${\cal R} = N^{1/2}$).
This fact validates the assumption ${\cal R} = N^{1/2}$ 
made in the monodisperse theory (see Section 3).
In fact, the fractal dimension close to 2 is a general consequence 
of dust growth without collisional compaction 
when aggregate collision is driven by Brownian motion and differential drift \citepalias{OTS09}.
Detailed inspection shows that values $D= 1.95$ and $2.03$ better fit to the data
if the fitted region is limited to the Brownian motion regime ($N<N_D$) and the differential drift regime ($N>N_D$), respectively.
The differential drift leads to a slightly higher fractal dimension than Brownian motion 
because the former reduces the collision rate for similar-sized aggregates 
(see Figure 15 of \citetalias{OTS09}).

\subsection{With Charging}
 \begin{figure}
\plotone{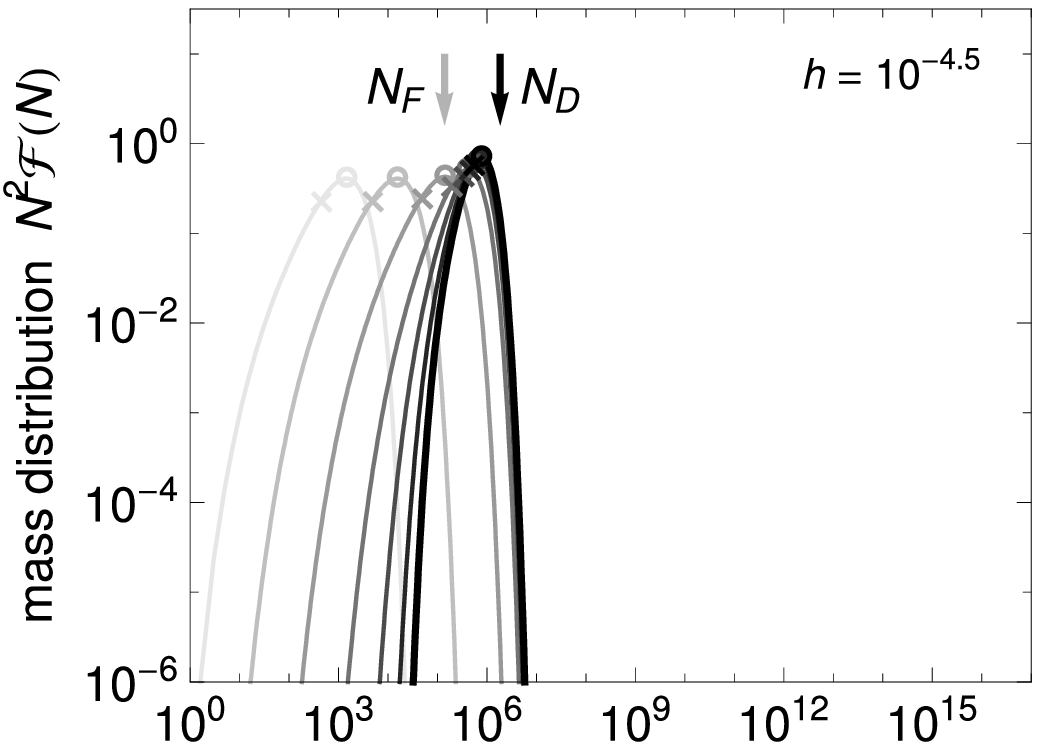}
\vspace{1mm}
\plotone{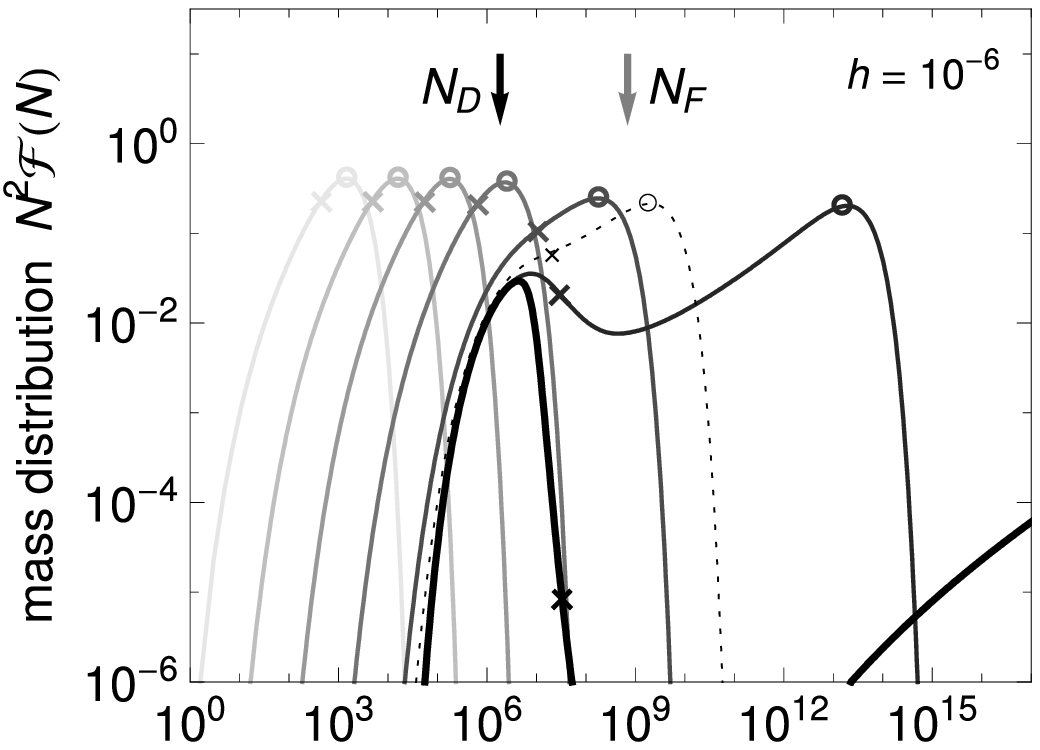}
\vspace{1mm}
 \plotone{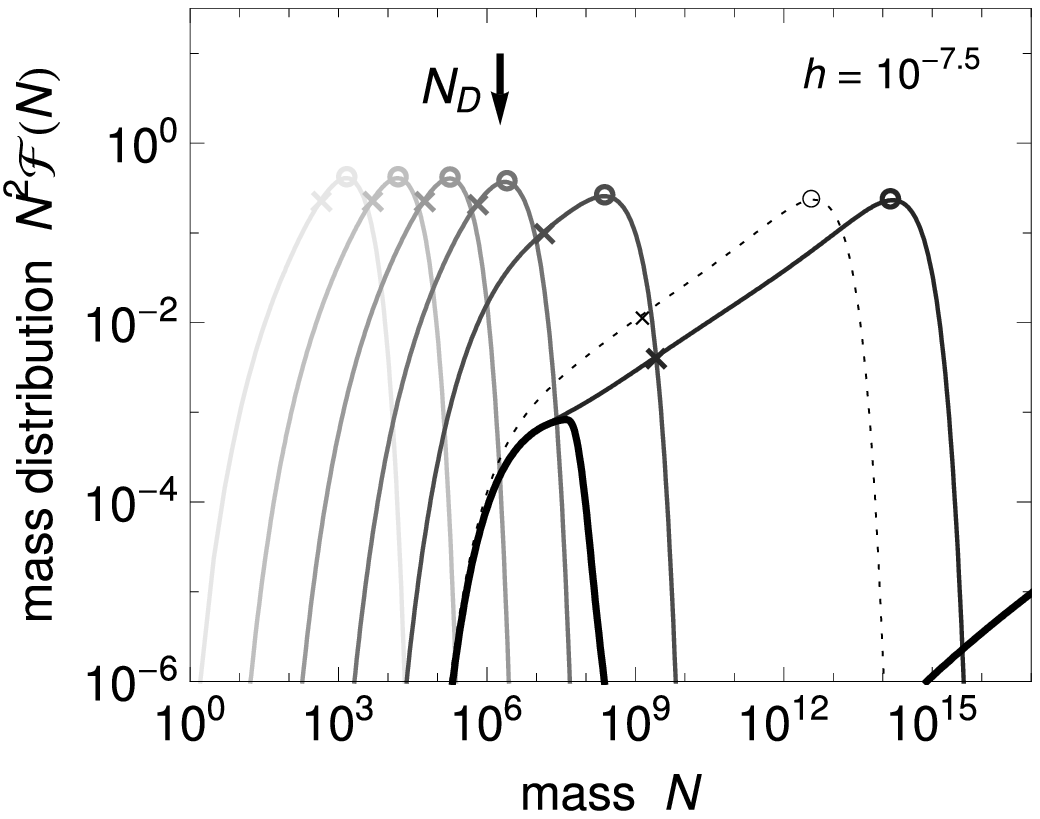}
 \caption{
Same as the upper panel of Figure~\ref{fig:evol0}, but for three charged cases, 
$h = 10^{-4.5}$, $10^{-6}$, and $10^{-7.5}$ (from top to bottom).
The other parameters are set to $(f_D\eps^2, f_E, \eps, \Psi) = (10^{-7},10, 10^{-1}, 10^{0.5})$.
The gray arrows indicate the freezeout mass $N_F$ predicted from the monodisperse theory.
The dotted curves in the middle and bottom panels 
show the mass distribution when the surface potential $\Psi$
exceeds the critical value $\Psi_\star$ (Equation~\eqref{eq:Psi_star}).
 }
 \label{fig:evol}
 \end{figure}

Now we show how the charging alters the evolution of the size distribution.
As in Section~3, we consider three cases of $h = 10^{-4.5}$, $10^{-6}$, and $10^{-7.5}$
with ($f_D\eps^2$, $f_E$, $\eps$) = ($10^{-7}$, $10$, $10^{-1}$).

In Figure~\ref{fig:evol}, we show the temporal evolution of the mass distribution ${\cal F}(N)$ 
for the three cases. The mass--radius relation ${\cal R}(N)$ is not shown here
because it is very similar to that for the uncharged case.
For $h=10^{-4.5}$, the monodisperse theory gives $({\cal E}_E/{\cal E}_K)_{\rm max} >1$, 
predicting the freezeout of the growth at $N \approx N_F \approx 10^{5.2}$ (see Table~\ref{table1}).
As expected, the evolution of the mass distribution starts to slow down at $N \approx N_F$, 
ending up with nearly monodisperse distribution peaked at $N \approx 10^6$.
In the simulation, we have followed the evolution at ${\cal T} = 10^6$, 
but observed no significant growth after ${\cal T}>10^4$.
 
For $h =10^{-6}$ and $10^{-7.5}$, by contrast, 
the outcome is qualitatively different from the prediction by the monodisperse theory,
as is shown in the middle and bottom panels of Figure~\ref{fig:evol}, respectively.
For the case of $h =10^{-6}$, the prediction was that the freezeout occurs 
at $N \approx N_F \approx 10^9$.
However, the simulation shows the size distribution evolving into a {\it bimodal} distribution,
in which one peak stays at $N \approx N_D$ and the other continues growing towards larger $N$.
Interestingly, similar behavior is seen in the case of $h = 10^{-7.5}$ 
despite the fact that the charging did not affect dust growth for this case
within the monodisperse theory.

 \begin{figure}
\plotone{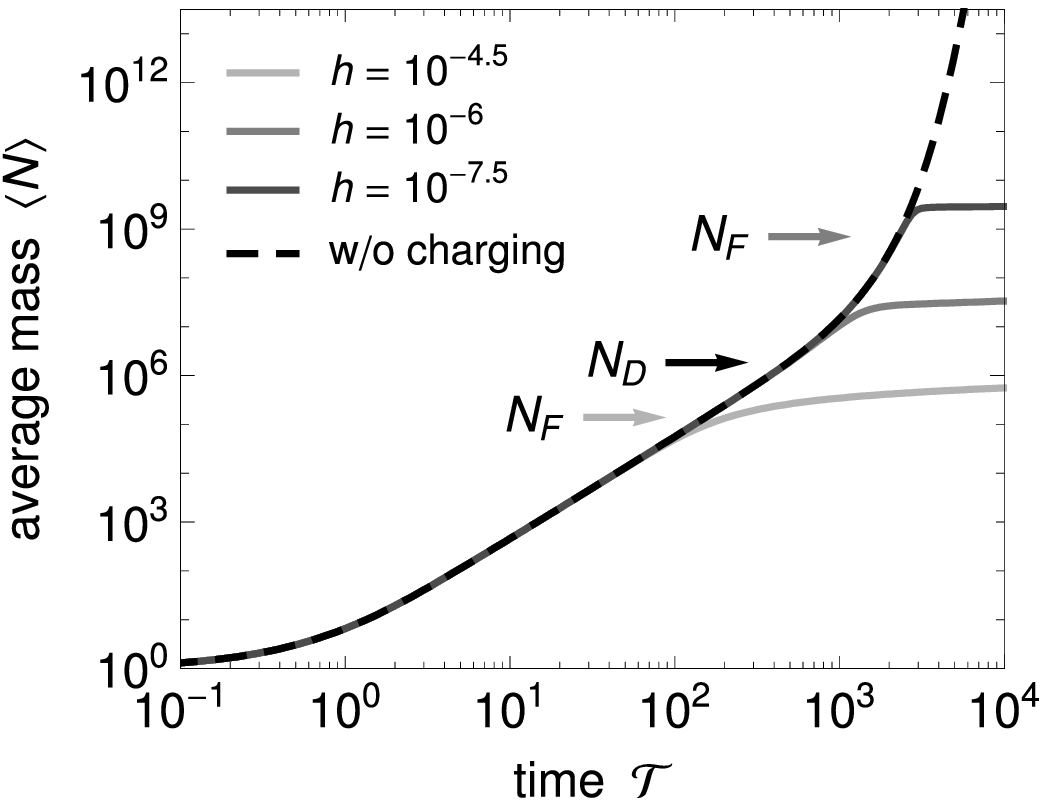}
\vspace{2mm}
\plotone{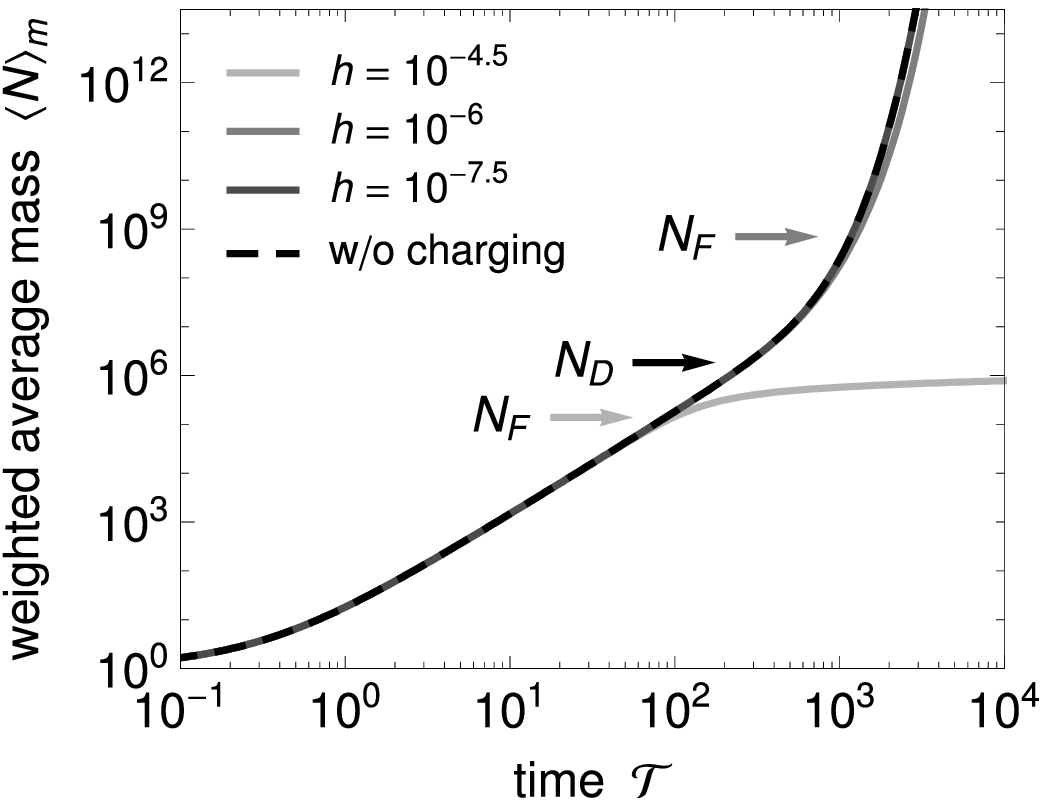}
 \caption{
Evolution of the average mass $\bracket{N}$ (upper panel) and the weighted average mass 
$\bracket{N}_m$ (lower panel) as a function of time ${\cal T}$.
The gray curves indicate the results for three charged cases of $h = 10^{-4.5}$, $10^{-6}$, $10^{-7}$,
while the black dashed curves are for the uncharged case ($h = 0$).
The other parameters are set to $(f_D,f_E,\eps,\Psi_\infty) = (10^{-5},10,10^{-1},10^{0.5})$.
The gray and black arrows indicate the critical drift mass $N_D$ and the freezeout mass $N_F$ 
predicted by the monodisperse theory, respectively.
}
 \label{fig:tN}
 \end{figure}
The evolution of the size dispersion can be better understood 
if we look at the evolution of $\bracket{N}$ and $\bracket{N}_m$.
Figure~\ref{fig:tN} compares them  
among the three charged cases together with the uncharged case.
See also Figure~\ref{fig:MDT_tN} in which the prediction from the monodisperse theory is shown.
For $h = 10^{-4.5}$, both $\bracket{N}$ and $\bracket{N}_m$ evolves 
as the monodisperse theory predicts.
However, for $h = 10^{-6}$ and $10^{-7.5}$, $\bracket{N}$ stops growing at certain values,
while $\bracket{N}_m$ continues growing as for the uncharged case.
This means that, in the latter cases, only a small number of aggregates continue growing
but nevertheless carry the greater part of dust mass in the system.

As we explain below, the transition to the bimodal distribution can be characterized by three steps:
\begin{enumerate}
\item At $\bracket{N}_m > N_D$, a long tail is formed at the low-mass end of the size distribution. 

\item Since aggregates belonging to the low-mass tail have a relatively small kinetic energy, 
they stop growing as the surface potential $\Psi$ reaches a certain value $\Psi_\star$
(see Equation~\eqref{eq:Psi_star} below). 
These ``frozen'' aggregates provide the total capacitance ${\cal C}_{\rm tot}$
which no longer decreases with time.
This leads to {\it the surface potential $\Psi$ of all aggregates no longer increasing with time}.

\item Consequently, aggregates of higher mass
 are less charged than in the case of the monodisperse growth.
The growth of the high-mass aggregates is no longer 
inhibited by the charge barrier.
\end{enumerate}
The first step was already discussed in the previous subsection.
Here, we explain how the second step follows after the development of the low-mass tail.
Let us approximate the mass distribution at the end of the first stage into two subgroups, 
one representing the high-mass side and the other representing the low-mass tail.
We characterize them with masses $N_1\gg N_D$ and $N_2 \approx N_D$.
The number of the low-mass aggregates decreases through their mutual collisions (``2--2 collision'')
and through sweep-up by the high-mass aggregates (``1--2 collision'').
This leads to the decrease in the total capacitance ${\cal C}_{\rm tot}$ and, in turn,
the increase in the surface potential $\Psi$.
We now write the relative kinetic energies for 1--2 and 2--2 collisions 
as ${\cal E}_{K,12}$ and ${\cal E}_{K,22}$.
Using Equations~\eqref{eq:ED}, \eqref{eq:EK}, and \eqref{eq:ND} together with 
$N/{\cal A} \approx b$ and $N_1 \gg N_2 \approx N_D$, these energies are approximately
evaluated as ${\cal E}_{K,12} \approx  1+2N_2/N_D \approx 3$ and 
 ${\cal E}_{K,22} \approx 1+N_2/N_D \approx 2$, respectively.
Note that ${\cal E}_{K,12}$ is nearly independent of $N_1$ because the reduced mass is determined by
smaller aggregates and because the drift velocity $\propto N_1/{\cal A}_1$ is nearly constant 
at large $N_1$.
Meanwhile, the electrostatic energies (Equation~\eqref{eq:EE}) for 1--2  and 2--2 collisions  are written as
  ${\cal E}_{E,12} \approx f_E(\Psi/\Psi_\infty)^2{\cal R}_2 \approx  f_E(\Psi/\Psi_\infty)^2N_D^{1/2}$ and
 ${\cal E}_{E,22} \approx f_E(\Psi/\Psi_\infty)^2{\cal R}_2/2 \approx  f_E(\Psi/\Psi_\infty)^2N_D^{1/2}/2$, respectively.
Again, ${\cal E}_{E,12}$ is independent of $N_1$, because the reduced radius 
is  determined by smaller aggregates.
Thus, the energy ratios for 1--2 and 2--2 collisions are obtained as
\beq
\frac{{\cal E}_{E,12}}{{\cal E}_{K,12}} \approx \frac{f_E \Psi^2 N_D^{1/2}}{3\Psi_\infty^2},
\qquad  \frac{{\cal E}_{E,22}}{{\cal E}_{K,22}} \approx   \frac{f_E \Psi^2 N_D^{1/2}}{4\Psi_\infty^2},
\label{eq:Eratio1222}
\eeq
independently of $N_1$.
Both the energy ratios exceed unity when $\Psi \ga \Psi_\star$,
where
\beq
\Psi_\star \equiv \left(\frac{4}{f_E N_D^{1/2}}\right)^{1/2}\Psi_\infty
 \approx  2\left( \frac{b f_D^{1/2}\eps}{f_E}\right)^{1/2} \Psi_\infty.
\label{eq:Psi_star}
\eeq
Note that $\Psi_\star$ is independent of $h$.
For $f_D\eps^2 = 10^{-7}$ and $f_E = 10$,
we obtain $\Psi_\star \approx 0.02\Psi_\infty$.

\begin{figure}
\plotone{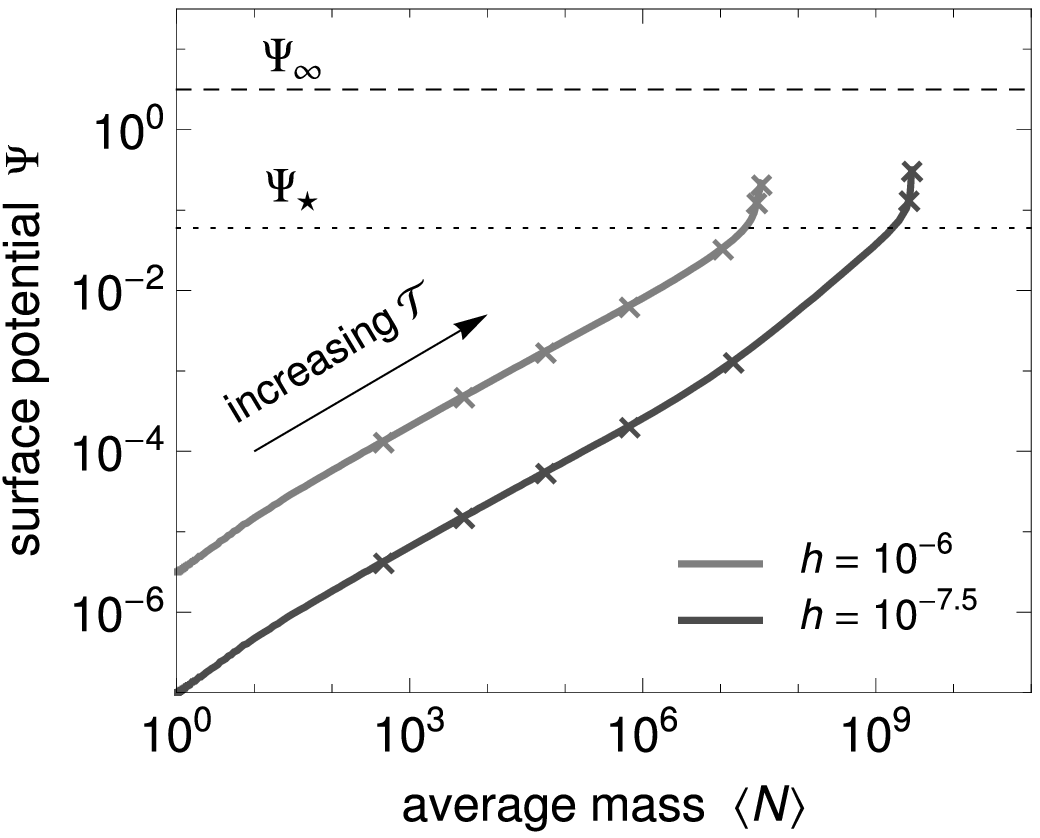}
\vspace{2mm}
\plotone{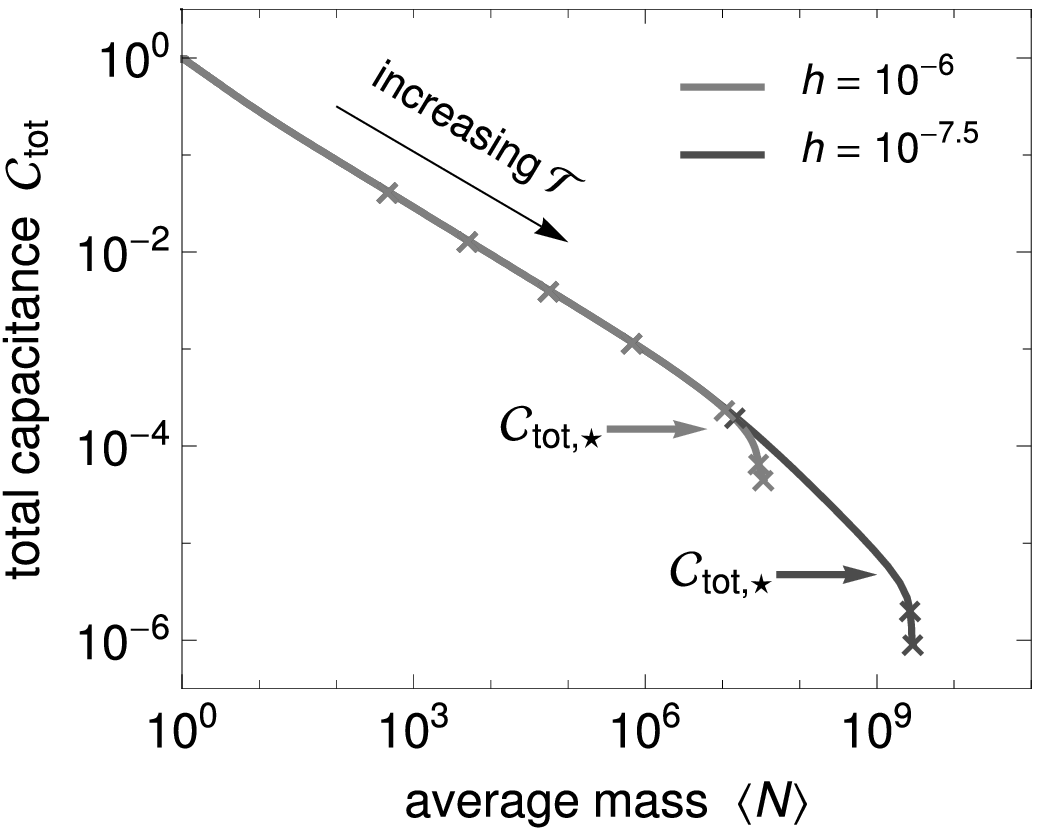}
\caption{
Surface potential $\Psi$ (upper panel) and total capacitance (lower panel) 
for $h =10^{-6}$ and $10^{-7.5}$ as a function of the average mass $\bracket{N}$. 
The dashed and dotted lines show $\Psi_\infty$ 
and $\Psi_\star$ (Equation~\eqref{eq:Psi_star}), respectively. 
The cross symbols indicate the values at ${\cal T} = 10^1$, $10^{1.5}$, 
$10^2, \dots,10^{4}$ (bottom to top).
}
 \label{fig:PsiCtot}
 \end{figure}

The above consideration suggests that the freezeout of the low-mass aggregates
occurs when $\Psi$ exceeds the critical value $\Psi_\star$.
To confirm this, in the upper panel of Figure~\ref{fig:PsiCtot}, 
we plot $\Psi$ versus the average mass $\bracket{N}$ for $h = 10^{-6}$ and $10^{-7.5}$.
We see that the increase in $\bracket{N}$ stops when $\Psi$ exceeds $\Psi_\star$.

It should be noted that the evolution of $\Psi$ is also slowed down for $\Psi\ga \Psi_\star$.   
This is because the ``frozen'' small aggregates govern the total electric capacitance 
${\cal C}_{\rm tot}$ of the system.
Using $\Psi \approx \Theta \approx b h\Psi_\infty/ {\cal C}_{\rm tot}$ (as is for the IDP limit), 
the total capacitance when $\Psi \approx \Psi_\star$ can be evaluated as
\beq
{\cal C}_{\rm tot} \approx {\cal C}_{\rm tot,\star} \equiv \frac{bh\Psi_\infty}{\Psi_\star} 
\approx \frac{(b f_E)^{1/2}h}{2(f_D\eps^2)^{1/4}}.
\label{eq:Ctot_star}
\eeq
The values of ${\cal C}_{\rm tot,\star}$ for the two cases are indicated 
in the lower right panel of Figure~\ref{fig:PsiCtot}.

We are now able to explain why the high-mass aggregates can grow beyond 
$N\approx N_F$ in the case of $h = 10^{-6}$.
First note that they can grow only through their mutual collisions (``1--1 collision'')
because 1--2 collisions have been already inhibited.
The relative kinetic energy and electrostatic energy for 1--1 collisions
are now given by ${\cal E}_{K,11} \approx 1 + N_1/N_D$ 
and ${\cal E}_{E,11} \approx (f_E/2) (\Psi_\star/\Psi_\infty)^2 N_1^{1/2}$.
Using $N_1 \gg N_D$ and Equation~\eqref{eq:Psi_star},
we obtain
\beq
\frac{{\cal E}_{E,11}}{{\cal E}_{K,11}} 
\approx \frac{f_E\Psi_\star^2 N_D}{2 \Psi_\infty^2 N_1^{1/2}
} 
\approx 2\pfrac{N_D}{N_1}^{1/2} \ll 1.
\label{eq:Eratio11}
\eeq
Thus, we find that the energy ratio decreases with mass, 
and therefore the growth of the high-mass aggregates is no longer inhibited by the charge barrier.
This is essentially due to the frozen aggregates keeping the surface potential $\Psi$ nearly constant. 
Without the frozen aggregates, $\Psi$ would increase as $N_1^{1/2}$, 
and the electrostatic energy ${\cal E}_{E,11}\propto N_1^{3/2}$
would take over ${\cal E}_{K,11} \propto N_1$
at a certain mass as in the monodisperse case.
With the frozen aggregates, by contrast, ${\cal E}_{E,11}$ increases only as $N_1^{1/2}$,
so cannot exceed ${\cal E}_{K,11}$.
It should be noted that the increasing kinetic energy will in reality 
cause collisional compaction at some stage, but this effect is neglected in our calculation.

One might wonder why the freezeout of the entire mass distribution occurs 
for $h = 10^{-4.5}$.
The key difference between the two cases $h = 10^{-4.5}$ and $h = 10^{-6}$
is the timing at which the electrostatic barrier becomes effective.
In the former case, the charge barrier becomes effective 
when the relative motion between aggregates is dominated by Brownian motion (i.e., $N_F< N_D$).
In this case, the aggregates cannot overcome the barrier even if $\Psi$ is kept constant,
since the electrostatic energy ${\cal E}_E \propto \Psi^2 N^{1/2}$  grows with mass
while the kinetic energy ${\cal E}_K \approx 1$ does not.
In the latter case, by contrast, the charge barrier becomes effective after 
the relative motion has been already dominated by the differential drift  (i.e., $N_D < N_F$).
In this case, the kinetic energy ${\cal E}_K \propto N$ can surpass the electrostatic energy
if $\Psi$ is kept constant.

Finally, we remark that $\Psi_\star$ can exceed $\Psi_\infty$ when $f_D\eps^2/f_E^2$ 
is sufficiently large (see Equation~\eqref{eq:Psi_star}).
In reality, however, the surface potential does not grow larger than $\Psi_\infty$.
For such cases, the energy ratios in Equation~\eqref{eq:Eratio1222} 
never exceed unity, so we expect that  
low-mass aggregates do not stop growing.
We will confirm this expectation in the following subsection.

\subsection{The Growth Criteria}
\begin{figure*}
\plottwo{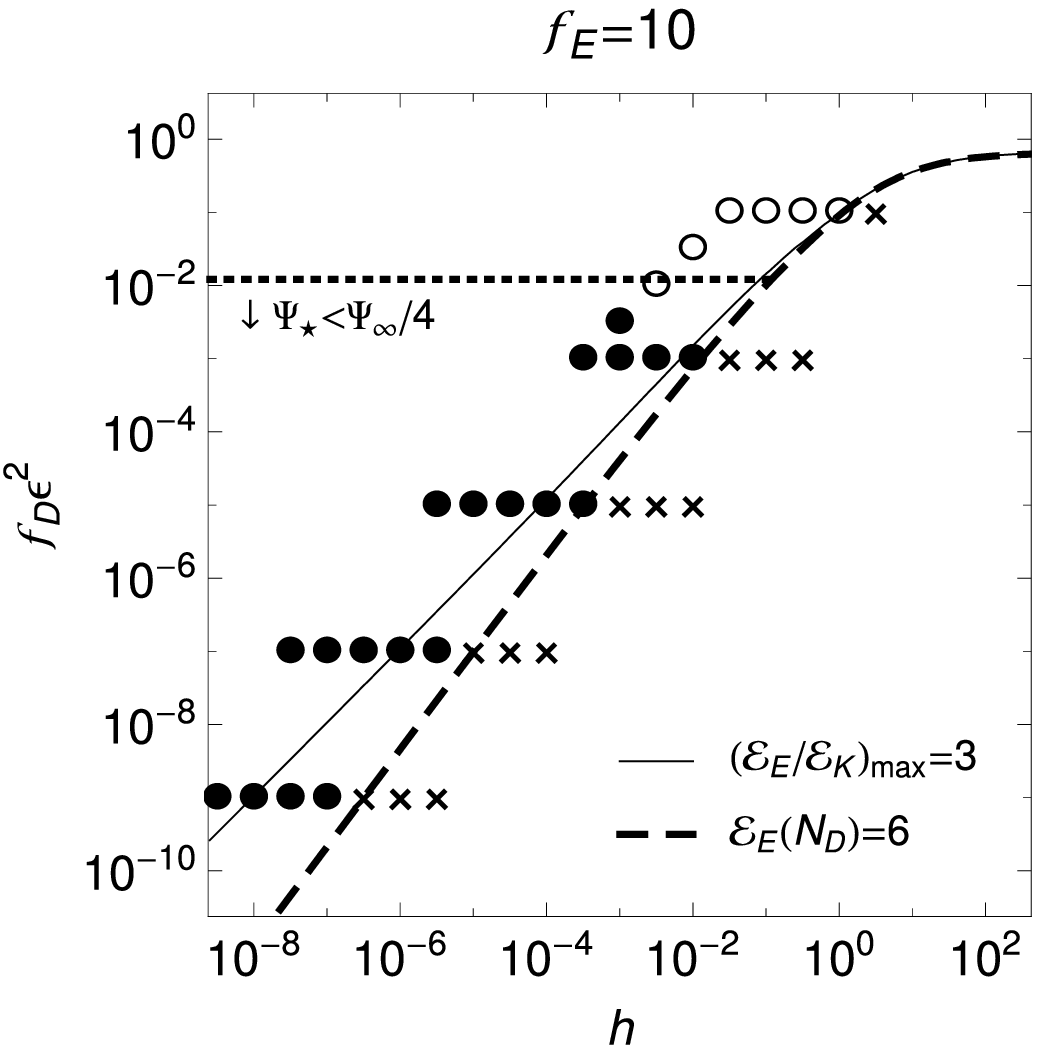}{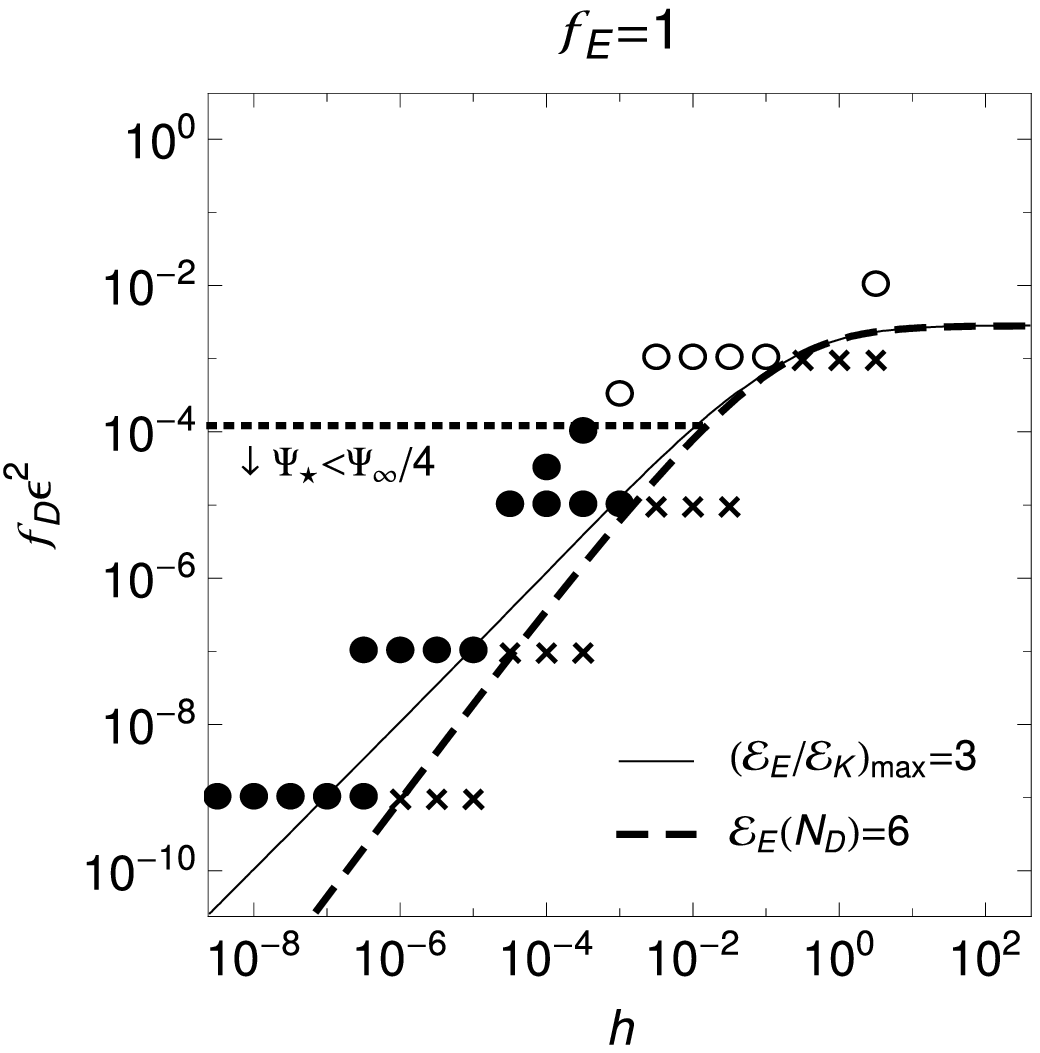}
\caption{
Outcome of numerical simulations for various parameters.
The crosses ($\times$) show the parameters for which
both $\bracket{N}$ and $\bracket{N}_m$ stop growing at $N\approx N_F$
(``total freezeout'').
The filled circles ($\bullet$) indicate the parameters for which $\bracket{N}$
stops growing at $N \approx N_D$ while $\bracket{N}_m$ does not (``bimodal growth'').
The open circies ($\circ$) indicate where both $\bracket{N}_m$ and $\bracket{N}$ continue growing
with a single-peaked distribution (``unimodal growth'').
The solid, dashed, and dotted lines show where 
 $({\cal E}_{E}/{\cal E}_{K})_{\rm max}=3$, ${\cal E}_{E}(N_D)=6$, 
 and $\Psi_\star = \Psi_\infty/4$, respectively.
 }
\label{fig:range}
\end{figure*}
The above examples suggest that the criterion $({\cal E}_E/{\cal E}_K)_{\rm max} \la 1$
for the monodisperse growth no longer applies 
when the evolution of the size distribution is taken into account.
To obtain a working criterion, we have performed numerical simulations 
for various sets of parameters $(f_D \eps^2,f_E,h)$. 

Figure~\ref{fig:range} shows the parameter space considered in the simulations.
We have chosen various sets of parameters ($f_D\eps^2$, $f_E$, $h$)
for which $({\cal E}_E/{\cal E}_K)_{\rm max} $ falls within the range $0.1 \dots 10^3$.
We have set $\eps = 10^{-1}$ in all of the simulations.

We find that the outcome of dust evolution can be classified into three types
in terms of the temporal evolution of $\bracket{N}$ and $\bracket{N}_m$.
In the first type, we observe that both $\bracket{N}$ and $\bracket{N}_m$ 
stop growing at $N\approx N_F$.
The outcome is characterized by frozen aggregates with a nearly monodisperse
distribution peaked at $N \approx N_F$ as seen in the top panel of Figure~\ref{fig:evol}.
 as seen in the top panel of Figure~\ref{fig:evol}.
We will refer to this type of growth outcome as the {\it total freezeout}.
In the second type, we see that $\bracket{N}$ stops growing at a certain value while 
$\bracket{N}_m$ continues growing.
The outcome is a double-peaked size distribution consisting of 
low-mass aggregates frozen at $N \approx N_D$ and ever-growing high-mass aggregates,
as seen in the middle and bottom panels of Figure~\ref{fig:evol}. 
We will call this type the {\it bimodal growth}.
In the third type,  we observe that both $\bracket{N}$ and $\bracket{N}_m$ continue growing.
The outcome is a single-peaked distribution of ever-growing aggregates
as is for uncharged cases (see Figure~\ref{fig:evol0}).
We will call this type the {\it unimodal growth}
to emphasize that the size distribution is characterized by a single peak.

The outcome of the growth for each set of parameters is displayed in Figure~\ref{fig:range}.
Here, the crosses ($\times$), filled circles  ($\bullet$), and open circles ($\circ$)
show the parameter sets for which we have observed the total freezeout, 
bimodal growth, and unimodal growth, respectively.
It is seen that the total freezeout occurs for small $f_D\eps^2$ and large $h$,
while the unimodal growth occurs when $f_D\eps^2$ is small.
 
First, we examine whether the total freezeout regime 
can be well represented by a criterion of the form 
$({{\cal E}_E}/{{\cal E}_K})_{\rm max} > {\rm constant}$
as suggested by the monodisperse theory (see Equation~\eqref{eq:growthcond_MDT}).
In Figure~\ref{fig:range}, we show a criterion $({{\cal E}_E}/{{\cal E}_K})_{\rm max} > 3$ 
with the solid curve.
It is seen that this criterion applies well at large $f_D\eps^2$ 
while it overestimates the size of the freezeout region at smaller $f_D\eps^2$.
It is clear that such a type of criteria do not explain the condition for the total freezeout to occur.
 
However, a criterion applicable for all parameter ranges
can be obtained if we slightly modify Equation~\eqref{eq:growthcond_MDT}.
The point is that the total freezeout is observed only in the Brownian motion regime,
i.e., only when the freezeout mass $N_F$ is smaller than the drift energy $N_D$.
This fact suggests that the total freezeout does not occur
if $({{\cal E}_E}/{{\cal E}_K})_{\rm max} \ga 1$ {\it but} ${\cal E}_E(N_D) \ll 1$
(this is the case for the parameter region (iii) in Figures \ref{fig:NF} and \ref{fig:EKEE_NF}).
This expectation motivates us to introduce another energy ratio,
\beq 
\frac{{\cal E}_E(N_D)}{{\cal E}_K(N_D)}
= \frac{{\cal E}_E(N_D)}{2},
\eeq
where we have used the definition of $N_D$, i.e., ${\cal E}_K(N_D) = 2$.
Note that ${\cal E}_E(N_D)/2$ is the maximum value of ${\cal E}_E/{\cal E}_K$ in
the Brownian motion regime 
because ${\cal E}_E$ monotonically increases with $N$ and ${\cal E}_K \leq 2$
at $N \leq N_D$.
In Figure~\ref{fig:range}, we show the line ${{\cal E}_E}(N_D) = 6$ with the dashed curve.
We see that the line represents the boundary of the total freezeout regime very well.
Thus, we conclude that the criterion for the total freezeout to occur is given by
\beq
{\cal E}_E(N_D) \ga 6.
\label{eq:growthcond}
\eeq

A simple criterion is also obtained for the 
boundary between the bimodal and unimodal growth regimes. 
As mentioned in Section 4.2, the bimodal growth occurs only if the critical surface potential
$\Psi_\star$ (Equation~\eqref{eq:Psi_star}) is lower than $\Psi_\infty$.
In Figure~\ref{fig:range}, we show the line $\Psi_\star = \Psi_\infty/4$ with the dotted curve. 
with the dashed curve.
We find that the condition for the bimodal growth to occur 
instead of the unimodal growth is given by
\beq
\Psi_\star \la \frac{\Psi_\infty}{4}.
\label{eq:growthcond2}
\eeq

To summarize, the outcome of charged dust growth can be classified 
into three cases (Table~\ref{table2}).
If ${\cal E}_E(N_D) \ga 6$, all aggregates stops growing before 
the systematic drift dominates their relative velocities.
The outcome is a nearly monodisperse distribution of frozen aggregates 
with typical mass $\approx N_F$.
If ${\cal E}_E(N_D) \la 6$ and $\Psi_\star \la \Psi_\infty/4$,
a large number of aggregates stop growing, 
but the major part of dust mass within the system is carried by a small number 
of ever-growing aggregates.
If ${\cal E}_E(N_D) \la 6$ and $\Psi_\star \ga \Psi_\infty /4$, 
all aggregates continue growing with a single-peaked size distribution.
The second case includes situations 
where no aggregates could continue growing
if the size distribution is limited to a monodisperse one.
This means that size distribution must be taken into account 
when we discuss how the charging of aggregates affects their collisional growth.

\begin{deluxetable}{ccl}
\tablecaption{Three Outcomes of the Growth of Charged Dust}
\tablecolumns{3}
\tablehead{
\multicolumn{2}{c}{Conditions} & \colhead{Outcome}
}
\startdata
${\cal E}_E(N_D) \ga 6$ & \nodata & Total freezeout \\
${\cal E}_E(N_D) \la 6$ & $\Psi_\star \la \Psi_\infty/4$ &  Bimodal growth  \\
${\cal E}_E(N_D) \la 6$ & $\Psi_\star \ga \Psi_\infty/4$ & Unimodal growth
\enddata
\label{table2}
\end{deluxetable}

\section{Discussion}
\subsection{An Application to a Protoplanetary Disk Model}
The growth criteria derived in Section~4 are general in a sense 
that no protoplanetary disk model is specified.
Although application to particular disk models is the subject of Paper II,
we will show here one example of how to use the criteria.

Here, we adopt the minimum-mass solar nebular (MMSN) model of \citep{Hayashi81}.
In this model, the gas temperature $T$ and the Kepler rotational frequency $\Omega_{\rm K}$
are given by $T = 280(r/1~\AU)^{-1/2}~{\rm K}$ and 
$\Omega_{\rm K} = (2\pi/1~{\rm yr})(r/1~\AU)^{-3/2}~{\rm rad~s^{-1}}$, 
where $r$ is the distance from the Sun.
The gas density $\rho_g$ and the vertical component of the stellar gravity $g$ 
are given by $\rho_g = 1.4\times 10^{-9} (r/1~\AU)^{-11/4}\exp(-z^2/2H^2)~{\rm g~cm^{-3}}$
and $g = \Omega_{\rm K}^2 z = 0.020 (r/1~\AU)^{-7/4}(z/H)$,
where $z$ is the distance from the midplane of the disk and 
$H = c_s/\Omega_{\rm K} = 5.0\times 10^{11}(r/1~\AU)^{5/4}~{\rm cm}$ is the gas scale height.
In this subsection, we neglect the effect of disk turbulence to dust collision 
and assume the stellar gravity as the only source of dust differential drift.
For the material density of monomers and the dust-to-gas mass ratio, 
we ignore the sublimation of ice for simplicity and set 
$\rho_0 = 1.4~{\rm g~cm^{-3}}$ and $\rho_d/\rho_g = 0.014$ \citep{THI05}. 
The maximum surface potential $\Psi_\infty$ is taken to be $2.81$
as is for $m_i = 24 m_{\rm H}$ and $s_i = 0.3$.
Substituting these relations into Equation~\eqref{eq:fD}, \eqref{eq:fE}, and $\eqref{eq:h}$ 
and setting $z=H$, we obtain
\beq
f_D = 4.1\times 10^{-5}\pfrac{a_0}{0.1~\micron}^5 \pfrac{r}{5~\AU}^{3},
\label{eq:fDr}
\eeq
\beq
f_E = 5.9 \pfrac{a_0}{0.1~\micron}\pfrac{r}{5~\AU}^{-1/2},
\label{eq:fEr}
\eeq
\beq
h = 2.0\times 10^{-3}\pfrac{a_0}{0.1~\micron}^3\pfrac{\zeta}{10^{-17}~{\rm s^{-1}}}
\pfrac{r}{5~\AU}^{7/2}.
\label{eq:hr}
\eeq
There equations give the radial profiles of ($f_D$, $f_E$, $h$) for the MMSN model 
at one scale height above the midplane.
In addition, we need to give the ionization rate $\zeta$ as a function of $r$.
Here, we simply give $\zeta = 10^{-17}$ at $r>3~\AU$ 
and $\zeta = 10^{-18}$ at $r<3~\AU$.
The higher value corresponds to ionization by cosmic rays and X-rays,
while the lower value corresponds to ionization by radionuclides.
The boundary $r = 3~\AU$ is chosen to approximate the full solution to $\zeta(r,z)$ 
including these ionizing sources \citepalias[see Figure~2(a) of][]{O09}.

\begin{figure}
\plotone{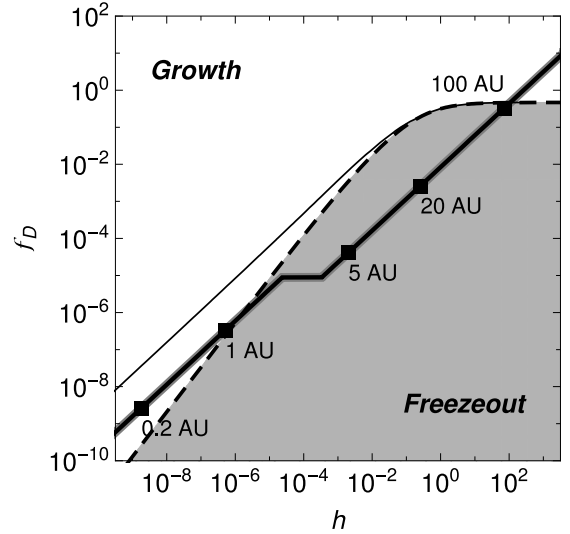}
 \caption{
Map of the minimum-mass solar nebular model in the $h$--$f_D$ plane.
The thick solid line shows the radial profile of $h$ (x-axis) and $f_D$ (y-axis)
at one scale height above the midplane of the disk,
with the filled squares indicating the distances from the central star. 
The break in the line approximates attenuation of cosmic-rays and X-rays at inner radii.
The gray region below the dashed curve indicates where 
we predict the total freezeout of fractal dust growth 
(see the freezeout condition, Equation~\eqref{eq:growthcond}).
Note that we have used a relation between $f_E$ and $f_D$ to project the freezeout region
onto in the $h$--$f_D$ plane (see text).
The thin solid line shows $({\cal E}_E/{\cal E}_K)_{\rm max} = 3$; 
fractal dust growth beyond the electrostatic barrier is possible between this line and the dashed line 
because of the effect of dust size distribution (see Section~4).
}
 \label{fig:MMSN}
 \end{figure}
Figure~\ref{fig:MMSN} illustrates how the MMSN model is mapped in the $h$--$f_D$ plane.
This thick solid line in the figure shows the radial profiles of $f_D$ and $h$ 
for $\eps=0.1$ and $a_0 = 0.1~\micron$.
The line moves upwards in the figure as $a_0$ is increased, because $f_D$ and $h$ are related as 
\beq
f_D = 8.4 \times 10^{-3}\pfrac{a_0}{0.1~\micron}^{17/7}
\pfrac{\zeta}{10^{-17}~{\rm s^{-1}}}^{-6/7}h^{6/7}
\label{eq:fDh}
\eeq
(this can be directly shown from Equations~\eqref{eq:fDr} and \eqref{eq:hr})
and hence $f_D$ increases with $a_0$ for fixed $h$.

Let us see the outcome of fractal dust growth in different locations of the disk
using the freezeout condition (Equation~\eqref{eq:growthcond}).
Since the condition depends on the three parameters ($f_D$, $f_E$, $h$),
the boundary between the growth and freezeout regions is a two-dimensional surface 
in the three-dimensional space.
However, it will be useful to represent the boundary  
as a single curve in the $h$--$f_D$ plane by relating $f_E$ to either $f_D$ or $h$.
Below, we use the relation $f_E = 1.1(a_0/0.1~\micron)^{11/6}f_D^{-1/6}$ 
obtained from Equations~\eqref{eq:fDr} and \eqref{eq:fEr}.

The thick dashed curve  in Figure~\ref{fig:MMSN} shows 
below which the freezeout condition holds for $a_0 = 0.1~\micron$ and $\eps=0.1$.
For this case, we see that the freezeout region covers 1--100 AU from the central star.
This means that the electrostatic barrier inhibits fractal dust growth except
in an inner region of $r\la 1~\AU$ and an very outer region of $r\ga 100~\AU$.
For comparison, we also show the line $({\cal E}_E/{\cal E}_K)_{\rm max} = 3$ 
with the thin solid curve (we again use the above relation between $f_E$ and $f_D$).
This line roughly corresponds to the boundary between the growth/freezeout regions 
predicted by the monodisperse theory (see Equation~\eqref{eq:growthcond_MDT}).
Comparing this line with the thick dashed curve, we see that
the inner region of $r\la 1~\AU$ would be also included in the freezeout region
if the bimodal growth mode as seen in Section 4 were not absent.
From this fact, we can expect that the bimodal growth is particularly important 
for dust evolution at small heliocentric distances. 
It should be noted, however, that all these results are dependent on 
the adopted disk model (e.g., laminar disk) and parameters (e.g., $a_0$).
We will defer further investigation to \citetalias{OTTS11b}.

\subsection{Effect of Charge Dispersion}
Up to here, we have assumed that all aggregates with the same radius 
have an equal charge $\bracket{Q}_a$.
In reality, the charge distribution has a nonzero variance, 
and hence aggregates can have a negative charge smaller than the mean value. 
Here, we show that the charge dispersion hardly affects the emergence of the total freezeout.

As shown in \citetalias{O09}, the charge distribution for aggregates of size $a$
is well approximated by a Gaussian distribution with variance (see Equation~(24) of \citetalias{O09})
\beq
\bracket{\delta Q^2}_a = \frac{1+\Psi}{2+\Psi}a \kB T.
\label{eq:Qvar}
\eeq
In principle, it is possible to fully take this effect into account  
by averaging the collision kernel $K$ over all $Q_1$ and $Q_2$. 
However, the average cannot be written in a simple analytic form.
For this reason, we simply estimate the effect of the charge dispersion as follows.
Clearly, the effect of the charge dispersion is significant only if 
$\bracket{\delta Q^2}_a$ is much larger than $\bracket{Q}_a^2$.
Using Equations~\eqref{eq:Q}, \eqref{eq:EE_def}, and~\eqref{eq:Qvar}, 
the ratio of $\bracket{\delta Q^2}_a$ to $\bracket{Q}_a^2$ can be written as
\beq
\frac{\bracket{\delta Q^2}_a }{\bracket{Q}_a^2} 
= \frac{1+\Psi}{2\bracket{{\cal E}_E}(2+\Psi)},
\label{eq:chargeratio}
\eeq
where $\bracket{{\cal E}_E}$ is the electrostatic energy for $Q_1 = Q_2 = \bracket{Q}_a$.
Since $1/2 \leq (1+\Psi)/(2+\Psi) \leq 1$ for all $\Psi$,
we find that the ratio $\bracket{\delta Q^2}_a/\bracket{Q}_a^2$ is of an order of 
$\bracket{{\cal E}_E}^{-1}$.
We also find that $\bracket{\delta Q^2}_a/\bracket{Q}_a^2$ {\it decreases} as dust grows
because $\bracket{{\cal E}_E}$ increases with $N$.

Using Equation~\eqref{eq:chargeratio}, let us consider 
whether the freezeout criterion (Equation~\eqref{eq:growthcond})
is affected by the presence of the charge dispersion.
With the charge dispersion ignored, 
the freezeout criterion is given by $\bracket{{\cal E}_E}(N_D) \ga 6$.
If this condition holds, we find from Equation~\eqref{eq:chargeratio}
that  $\bracket{\delta Q^2}_a(N_D) \la  0.08[\bracket{Q}_a(N_D)]^2(1+\Psi)/(2+\Psi) \la 0.08[\bracket{Q}_a(N_D)]^2$.
This means that the ``true'' value of ${\cal E}_E(N_D)$
(i.e., the value with the charge dispersion taken into account) 
is not much different from the ``approximate'' value $\bracket{{\cal E}_E}(N_D)$
as long as $\bracket{{\cal E}_E}(N_D) \ga 6$.
Hence,  the charge dispersion hardly affects the emergence of the total freezeout.

\subsection{Dependence on the Velocity Dispersion}
In this study, we have assumed that the velocity dispersion is thermal 
(see our probability distribution function, Equation~\eqref{eq:PDF}).
This assumption neglects any fluctuation in the drift acceleration $g$.
This will be reasonable if $g$ is caused by stellar gravity ($g = \Omega_{\rm K}^2z$).
By contrast, the validity of this approximation is unclear if $g$ is driven  
by turbulence ($g \approx u_\eta/t_\eta$).
For example, recent MHD simulations by \citet{CST08} suggest that 
$g$ may fluctuate by $10\%$ in MRI-driven turbulence.
To check the robustness of our conclusion,
we examine how the outcome of dust growth depends on 
the choice of the velocity dispersion.  

 \begin{figure}
\epsscale{1.1}
\plotone{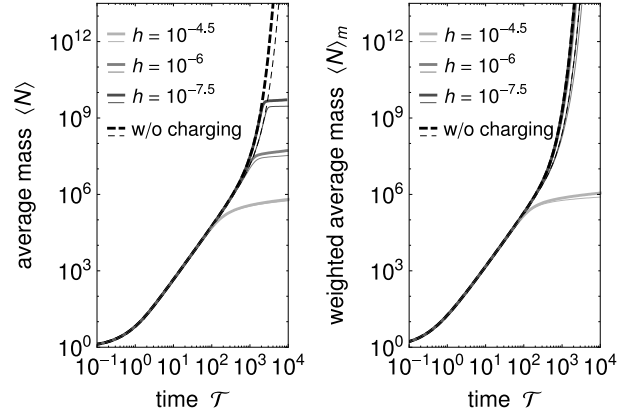}
 \caption{
Comparison of the temporal evolution of  $\bracket{N}$ (left panel) 
 $\bracket{N}_m$ (right panel) between different velocity dispersion models.
The thick curves are the results for a modified velocity dispersion model (see Section 5.2),
while the thin curves are the same as the curves showing in the upper panel of Figure~\ref{fig:tN}.
The thick and thin curves are very similar (indistinguishable for $h = 10^{-4.5}$ in the left panel), 
meaning that the dependence on the velocity dispersion is weak.
}
 \label{fig:tNturb}
 \end{figure}
Here, we consider the cases where the fluctuation in the differential drift velocity 
is as large as the mean value.
We mimic this situation by replacing $k_{\rm B}T$ in Equation~\eqref{eq:PDF} 
by $k_{\rm B}T + M_\mu (\Delta u_D)^2$, 
where $\Delta u_D$ is the mean relative velocity given by Equation~\eqref{eq:uD}.
With the modified velocity distribution function, 
we carried out simulations for four sets of parameters as in Section 4.2.
Figure~\ref{fig:tNturb} compare the evolution of $\bracket{N}$ and $\bracket{N}_m$  
obtained here with that in Section 4.2 (Figure~\ref{fig:tN}).
We find no significant difference between the two results.
This should be so since the freezeout occurs while Brownian motion dominates over 
the differential drift (i.e., $\kB T \gg M_\mu(\Delta u_D)^2$; see Section 4).
Detailed inspection shows that the average masses grow slightly faster 
when the dispersion is added to the differential drift, but this is clearly a minor effect.
Hence, we conclude that fluctuation in the differential drift velocity 
hardly affects the outcome of the dust growth.

\subsection{Validity of the Fractal Growth Model}
So far, we have assumed that dust grows into porous (fractal) aggregates.
This assumption is true only when the impact energy is so low 
that compaction of aggregates upon collision is negligible. 
Here, we show that the collisional compaction is actually negligible 
when we consider the freezeout of dust growth.

It has been shown by \citet{DT97} that the collisional compaction become effective 
when the impact energy exceeds $3E_{\rm roll}$, where
\beqn
E_{\rm roll} &=& 3\pi^2 \gamma a_0\xi_{\rm crit} \nonumber \\
&\approx& 6 \times 10^{-10} \pfrac{\gamma}{100{\rm~erg~cm^{-2}}}
\pfrac{\xi_{\rm crit}}{2{\rm\mathring{A}}}\pfrac{a_0}{0.1\micron} {\rm~erg} \qquad
\label{eq:Eroll}
\eeqn
is the energy needed for a monomer to roll on another monomer in contact by 90 degrees.
$\gamma$ is the surface energy per unit area and is estimated as $25{\rm~erg~cm^{-2}}$
for rocky monomers and somewhat higher for icy monomers.
$\xi_{\rm crit}$ is the critical rolling displacement for inelastic rolling and is
theoretically constrained as $>2{\rm\mathring{A}}$ \citep{DT95}.

As seen in the previous section, the total freezeout occurs 
when Brownian motion dominates aggregate collision. 
Hence, the relative kinetic energy between frozen aggregates is 
equal to the thermal energy $\sim \kB T$.
Assuming $T\sim 100{\rm~K}$, the thermal energy is $\sim 10^{-14}~{\rm erg}$,
which is many orders of magnitude lower than $E_{\rm roll}$.
Therefore, collisional compaction is negligible whenever the total freezeout occurs.

Of course, the compaction is no longer negligible when the electrostatic barrier is overcome
since the drift energy increases with aggregate mass and finally exceeds $E_{\rm roll}$.
Investigation of dust growth after the fractal growth stage is beyond the scope of this study.
 
\subsection{On the Role of Porosity Evolution}
As shown in the previous subsection, it is valid to assume the fractal dust growth
whenever we focus on the freezeout of dust growth. 
However, it has been still unclear whether the freezeout occurs even 
without the porosity evolution.
Indeed, in previous studies on dust coagulation, 
it is common to ignore the porosity evolution
and model aggregates as spheres of a fixed internal density  
\citep[e.g.,][]{W80,NNH81,THI05,BDH08}.
To fully understand the robustness of the freezeout,
we will discuss how the growth outcome changes if we adopt the compact aggregate model.

\subsubsection{Drift and Electrostatic Energies for Compact Dust Particles}
It is straightforward to write down the dimensionless 
energies ${\cal E}_D$ and ${\cal E}_E$ for the compact model.
Since ${\cal R} = N^{1/3}$ and ${\cal A} = N^{2/3}$ for compact particles,
Equations~\eqref{eq:ED} and \eqref{eq:EE} are now replaced by
\beq
{\cal E}_D = f_D \frac{N_1N_2}{N_1 + N_2} \left|\frac{N_1}{{\cal A}_1} -\frac{N_2}{{\cal A}_2} \right|^2
= f_D \frac{N_1N_2}{N_1 + N_2} \left|N_1^{1/3} -N_2^{1/3} \right|^2
\label{eq:ED_comp}
\eeq
and
\beq
{\cal E}_E = f_E \pfrac{\Psi}{\Psi_\infty}^2\frac{{\cal R}_1{\cal R}_2}{{\cal R}_1+{\cal R}_2}
= f_E \pfrac{\Psi}{\Psi_\infty}^2\frac{(N_1N_2)^{1/3}}{N_1^{1/3}+N_2^{1/3}},
\label{eq:EE_comp}
\eeq
respectively.
Note that $\eps$ identically vanishes here by the definition of the compact dust model.

\subsubsection{Simulations}
 \begin{figure}
\plotone{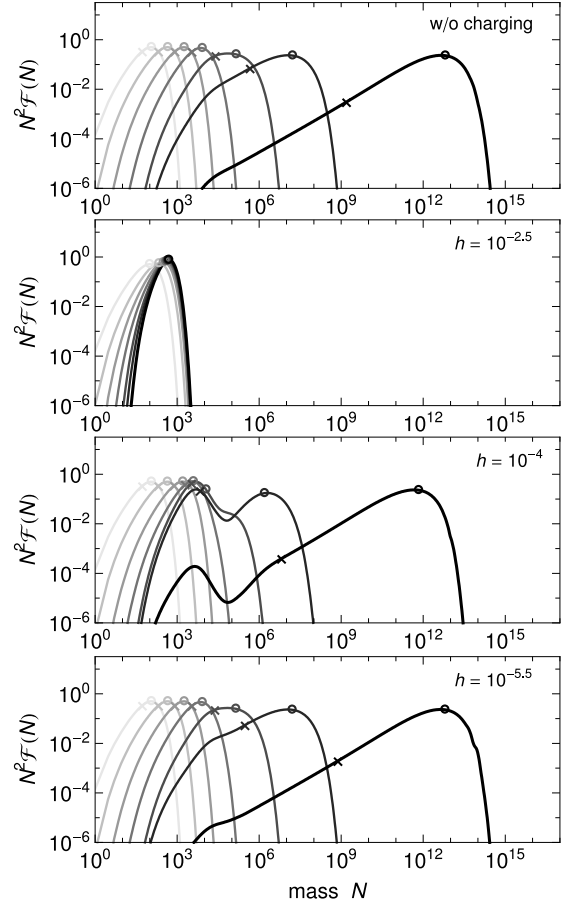}
 \caption{
Evolution of  the mass distribution function ${\cal F}(N)$ for {\it compact} dust models.
The gray curves show the snapshots of $N^2{\cal F}(N)$ at 
${\cal T} = 10^{1}$, $10^{1.5}$, $\dots,10^{3}$, $10^{3.3}$, $10^{3.7}$ (from left to right).
The crosses and open circles indicate $\bracket{N}$ and $\bracket{N}_m$ 
at different times, respectively.
Parameters $(f_D,f_E,\Psi_\infty)$ are set to $(10^{-5},10,10^{0.5})$.
}
 \label{fig:evolc}
 \end{figure}
Using Equations~\eqref{eq:ED_comp} and \eqref{eq:EE_comp} 
instead of Equations~\eqref{eq:ED} and \eqref{eq:EE},
we have carried out simulations for several sets of ($f_D$, $f_E$, $h$, $\Psi_\infty$).
Figure~\ref{fig:evolc} shows the results for the uncharged case ($h=0$)
and three charged cases ($h=10^{-2.5}$, $10^{-4}$, $10^{-5.5}$) with fixed 
$(f_D,f_E,\Psi_\infty) = (10^{-5},10,10^{0.5})$.
Note that the values of $(f_D,f_E,\Psi_\infty)$ are the same as those for the examples 
shown in Sections~4.1 and 4.2.

Without charging, the outcome of dust growth is qualitatively similar to that 
for the porous model (see the upper panel of Figure~\ref{fig:evol0}).
Namely, we see power-law growth at early times (${\cal T}\la 10^3$) 
and exponential growth at later times (${\cal T}\ga 10^3$). 
One important difference is that the exponential growth begins 
at a lower mass $N$ than in the porous case.
As already mentioned in Section 3, the exponential growth is an indication
that the differential drift takes over Brownian motion in the relative velocity between particles. 
In the porous model, the drift velocity of aggregates increases only slowly with mass,
because the fractal dimension is close to 2 and hence 
the mass-to-area ratio $N/{\cal A}$ is nearly insensitive to $N$.
In the compact case, by contrast, the drift velocity increases with $N$ 
($\Delta u_D \propto N/{\cal A} \propto N^{1/3}$).
For this reason, the drift motion takes over Brownian motion ($\Delta u \propto N^{-1/2}$)
at lower $N$ than in the porous case.   

The difference mentioned above consequently influences the outcome 
of dust growth with charging charging (the gray curves in Figure~\ref{fig:evolc}).
We see that the total freezeout does not occur at $h = 10^{-4}$ as it does in the porous case.
This is because of the faster increase in the differential drift velocity mentioned above.
In fact, the electrostatic energy also increases faster 
than in the compact case because of the faster decrease in 
the total projected area ${\cal A}_{\rm tot}$ and capacitance ${\cal C}_{\rm tot}$.
However, this effect is small compared to the faster increase in the kinetic energy. 
Therefore, we can say that the compact dust growth is resistive to the freezeout. 
Note that the compact growth is not free from the occurrence of the freezeout;
in fact, we observe the freezeout for a higher-$h$ case, $h = 10^{-2.5}$.

We see that the mass distribution for $h = 10^{-4}$ splits into two peaks.
However, the evolution is qualitatively different from what we call bimodal growth
in the porous case.
The difference is that the low-mass peak gets continuously depleted 
as the high-mass peak grows towards higher $N$.
This occurs because the high-mass particles acquire arbitrarily high 
drift velocities as they grow.
For the porous dust model, we have seen that the the impact energy 
for highly unequal-sized collisions, ${\cal E}_{K,12}$,
is nearly independent of the mass $N_1$ of the heavier particle (see Section~4.2).
In the compact model, by contrast, the impact energy is approximately given by 
${\cal E}_{K,12} \approx 1+ f_DN_2N_1^{2/3}$
 (which directly follows from Equation~\eqref{eq:ED_comp} with $N_1 \gg N_2$),
and this increases with $N_1$.
However, the electrostatic energy ${\cal E}_{E,12} \approx f_E(\Psi/\Psi_\infty)^2
{\cal R}_2$ is independent of $N_1$ as is in the porous case.
Hence, we find that a high-mass particle with sufficiently large $N_1$ 
can capture smaller particles\footnote{
Strictly speaking, the decrease in the number of low-mass particles leads to the increase in $\Psi$ 
(see Section 4.2), and hence proceeds in a way that ${\cal E}_{E,12}$ balances with
${\cal E}_{K,12}$ until $\Psi$ reaches $\Psi_\infty$.
}.

\subsubsection{Freezeout Criterion for the Compact Dust Model}
 \begin{figure}
\plotone{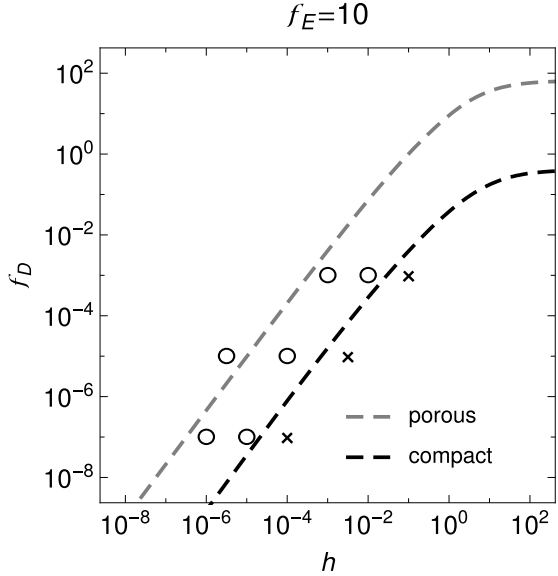}
 \caption{
Outcome of numerical simulations for {\it compact sphere} models
(see Figure~\ref{fig:range} for porous aggregate models).
The crosses ($\times$) show the parameters for which
both $\bracket{N}$ and $\bracket{N}_m$ stop growing at $N\approx N_F$
(``total freezeout''), while 
the open circies ($\circ$) indicate where both $\bracket{N}_m$ and $\bracket{N}$ 
continue growing (``unimodal growth'').
The black dashed curve shows the boundary below which ${\cal E}_{E}(N_D)$ exceeds $6$. 
For comparison, the boundary for porous models (the dashed curve in the left panel of Figure~\ref{fig:range})
is shown by the gray dashed curve.
}
 \label{fig:rangecomp}
 \end{figure}
Figure~\ref{fig:rangecomp} summarizes the results of the simulations for compact dust models.
The crosses and open circles indicate the sets of parameters for which we observe 
total freezeout and unimodal growth, respectively.
The gray dashed curve shows the boundary below which the freezeout condition satisfies
for the {\it porous} model, i.e., the black dashed curve in Figure\ref{fig:range}.
We see that the compact growth results in the freezeout in a more
restricted region of the parameter space than the porous growth.
It is clear that the compact model is less conducive to the freezeout
compared to the porous model.
  
To obtain a freezeout criterion for the compact model, 
it is useful to introduce ${\cal E}_D$ and ${\cal E}_E$ written 
as a function of a single mass $N$ rather than $N_1$ and $N_2$.
as done for the porous model.
There is no difficulty in evaluating ${\cal E}_E$ assuming that the particles are 
monodisperse, i.e.,  $N_1 = N_2 = N$.
Using ${\cal A}_{\rm tot} = {\cal A}/N = N^{-1/3}$ 
and ${\cal C}_{\rm tot} = {\cal C}/N = N^{-2/3}$, 
we have $\Theta = h\Psi_\infty N$.
Thus, the electrostatic energy for monodisperse compact particles is
\beq
{\cal E}_E = \frac{f_E}{2} \left[1+(hN)^{-0.8}\right]^{-2.5}N^{1/3}
\label{eq:EE1_comp}
\eeq
In contrast, we would obtain no meaningful expression for ${\cal E}_D$
within the exact monodisperse assumption because ${\cal E}_D$
identically vanishes for $N_1 = N_2 = N$.
For this reason, we will simply replace $N_1N_2/(N_1+N_2) \to N$ 
and $|N_1^{1/3}-N_2^{1/3}| \to N^{1/3}$ in Equation~\eqref{eq:ED_comp} to get 
\beq
{\cal E}_D = f_D  N^{5/3}.
\label{eq:ED1_comp}
\eeq
As done in Section~3.1, we can define the drift mass $N_D$ by ${\cal E}_D(N_D) = 1$;
using Equation~\eqref{eq:ED1_comp}, we have $N_D = f_D^{-3/5}$.
Hence, the critical energy ${\cal E}_E(N_D)$ for the compact model can be explicitly given 
as a function of $(f_D, f_E, h)$ by 
\beq
{\cal E}_E(N_D) = \frac{f_E}{2} \left[1+(hf_D^{-3/5})^{-0.8}\right]^{-2.5}f_D^{-1/5}.
\label{eq:EEND_comp}
\eeq 

Let us examine whether the condition for the freezeout is 
well described by the value of ${\cal E}_D(N_D)$ as is for the porous cases.
The black dashed curve in Figure~\ref{fig:rangecomp} 
shows the line where ${\cal E}_D(N_D)$ for the compact model  is equal to 6.
For comparison, the line ${\cal E}_D(N_D) =6$ for the porous case 
(i.e., the dashed curve in the left panel of Figure~\ref{fig:range}) is also shown 
by the gray dashed curve.
We find that the black line successfully explains 
the boundary between the freezeout and unimodal growth regions.
Hence, we conclude that the freezeout criterion for the compact model is
again given by Equation~\eqref{eq:growthcond} if only
we use Equation~\eqref{eq:EEND_comp} for ${\cal E}_D(N_D)$.

To summarize this subsection, we have investigated 
how the growth outcome changes if one adopts a compact dust model.
We confirmed that the total freezeout does occur even in the compact dust growth.
This means that a fractal dust model is {\it not} a prerequisite for the emergence of the freezeout.
However, this does not mean that the porosity evolution is negligible 
when we analyze the effect of electrostatic barrier against dust growth. 
As shown above, the compact model makes dust growth more resistive to the freezeout 
because the differential drift takes over Brownian motion at a lower mass.
Therefore, the porosity evolution must be properly taken into account 
in order not to overlook the significance of the electrostatic barrier. 

\section{Summary}
In this paper, we have investigated how the charging of dust affects its coagulation 
in weakly ionized protoplanetary disks.
In particular, we have focused on the effect of the dust size distribution,
which was ignored in the previous work \citepalias{O09}. 
We have used the porous (fractal) aggregate model recently proposed by \citetalias{OTS09}
to properly take into account the porosity evolution of aggregates.

To clarify the role of size distribution, we have divided our analysis into two steps. 
As the first step, in Section~3, we have presented a general analysis 
on the coagulation of charged aggregates under the monodisperse growth approximation. 
The monodisperse approximation allows us to define several useful quantities,
such as the maximum energy ratio $({\cal E}_K/{\cal E}_E)_{\rm max}$, 
the drift mass $N_D$, and the freezeout mass $N_F$.
We have shown that, if the maximum energy ratio $({\cal E}_K/{\cal E}_E)_{\rm max}$ is larger than unity, the monodisperse growth stalls (or "freezes out") at mass $N \approx N_F$,
as was predicted by \citetalias{O09}. 

As the second step, in Section~4, 
we have calculated dust coagulation using the extended Smoluchowski method \citepalias{OTS09}
to examine how the outcome changes when the size dispersion is allowed to freely evolve.
We find that, under certain conditions, the electrostatic repulsion 
leads to {\it bimodal} growth, rather than total freezeout.
This bimodal growth is characterized by a large number of  ``frozen'' aggregates
and a small number of ``unfrozen'' aggregates, 
the former controlling the charge state of the system
and the latter growing larger and larger carrying the major part of the system mass.

Based on the results of our numerical simulations, we have obtained a set of simple criteria 
that allows us to predict how the size distribution evolves for given conditions
(Section 4.3; Table~\ref{table2}). These read:
\begin{itemize}
\item
If ${\cal E}_E(N_D) \ga 6$, all aggregates stops growing before 
the systematic drift dominates their relative velocities ({\it total freezeout}).
The outcome is a nearly monodisperse distribution of frozen aggregates 
with typical mass $\approx N_F$.
\item
If ${\cal E}_E(N_D) \la 6$ and $\Psi_\star \la \Psi_\infty/4$,
a large number of aggregates stop growing, 
but the major part of dust mass within the system is carried by a small number 
of ever-growing aggregates ({\it bimodal growth}).
\item
If ${\cal E}_E(N_D) \la 6$ and $\Psi_\star \ga \Psi_\infty /4$, 
all aggregates continue growing with a single-peaked size distribution ({\it unimodal growth}).
\end{itemize}
The second case includes situations 
where aggregates cannot continue growing in the monodisperse growth model.
Thus, the size distribution is an important ingredient
for the growth of dust aggregates beyond the electrostatic barrier.

We emphasize again that our analysis assumed fractal evolution of dust aggregates.
This assumption is valid only when the collision energy 
is so small that collisional compaction is negligible \citep{DT97,SWT08}.
We have proven that the collisional compaction is indeed negligible 
as long as the total freezeout is concerned since the freezeout 
always occurs when Brownian motion dominates aggregate collision (Section 5.4). 
It should be noted that most theoretical studies on dust coagulation 
\citep[e.g.,][]{NNH81,THI05,BDH08} have ignored the porosity evolution
and modeled aggregates as compact spheres.
However, we have found that such simplification leads 
to underestimation of the electrostatic barrier
because compact spheres are frictionally less coupled 
to the gas and hence have higher drift velocities 
than porous aggregates of the same mass (Section 5.5).
Therefore, the porosity evolution must be properly taken into account 
when considering the electrostatic barrier against dust growth in protoplanetary disks.

In \citetalias{OTTS11b}, we apply our growth criteria 
to particular protoplanetary disk models to investigate 
the effect of the electrostatic barrier in the early stage of planet formation.
\acknowledgments
The authors thank the anonymous referee for the many comments 
that greatly helped improve the manuscript.
S.O. is supported by Grants-in-Aid for JSPS Fellows ($22\cdot 7006$) from MEXT of Japan.

\section*{Appendix \\Numerical Estimation of the Area Dispersion}
Let us consider two groups of porous aggregates each of which is characterized 
by aggregate mass $N_j(j=1,2)$.
In either group, aggregates have different values of the projected area $A_j$.
Therefore, the projected area, or the mass-to-area ratio $B_j \equiv N_j/A_j$,
of an aggregate randomly chosen from the $j$-th group can be regarded as a stochastic variable.  
The average of the quantity $|B_1-B_2|^2$ over all possible pairs is given by
\beqn
\ovl{|B_1-B_2|^2} &=& |\ovl{B}(N_1)-\ovl{B}(N_2)|^2
+ \sum_{j=1,2} \ovl{\delta B^2}(N_j)  \nonumber \\
&\equiv &  |\ovl{B}(N_1)-\ovl{B}(N_2)|^2 + \sum_{j=1,2}\eps(N_j)^2\ovl{B}(N_j)^2,
\label{eqA}
\eeqn
where $\ovl{B}(N_j)$ and $\ovl{\delta B^2}(N_j)$ are the statistical average and 
variance of $B$ for aggregates of the $j$-th group,
and $\eps(N) \equiv \ovl{\delta B^2}(N)^{1/2}/\ovl{B}(N)$.
Note that we have assumed that $B_1$ and  $B_2$
are uncorrelated, i.e., $\ovl{B_1 B_2} = \ovl{B}(N_1)\ovl{B}(N_2)$.
Equation~\eqref{eqA} reduces to Equation~\eqref{eq:N/A_eps}
 if $\eps(N)$ is independent of $N$. 
In this appendix, we estimate $\eps(N)$ using numerically created BCCA clusters.

\begin{figure}[t]
\plotone{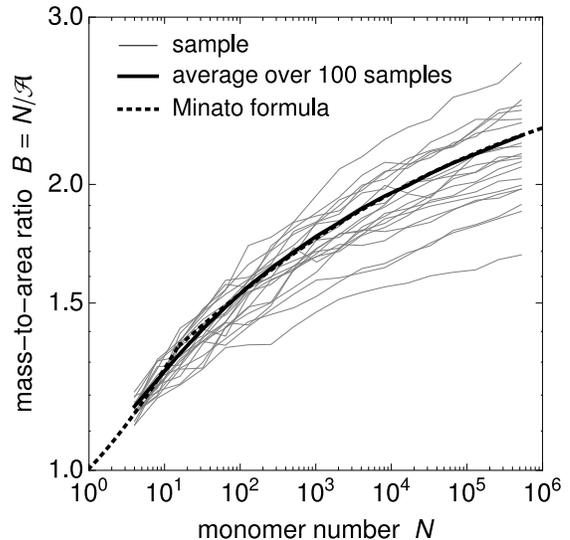}
\caption{
Mass-to-area-ratio $B=N/{\cal A}$ versus monomer number $N$ for numerically created 
BCCA clusters. The thin solid curves show 20 samples, while the thick solid curve
indicates the average over 100 samples.
The dashed curve shows Minato's formula (Equation~\eqref{eq:A_BCCA}).
}
\label{fig:MovA}
\end{figure}
 We have performed 100 BCCA simulations and obtained the relation between 
${\cal A}$ and $N$ for each run.
Since the projected area of an aggregate generally depends on the choice of the projection angle,
we determined it as the average over 15 randomly chosen orientations.
Figure~\ref{fig:MovA} shows the mass-to-area ratio $B$ versus monomer number $N$ 
for 20 samples as well as the average $\ovl{B}$ over 100 samples.
The area formula of \citet{Minato+06}, Equation~\eqref{eq:A_BCCA}, is also plotted 
to show that $\ovl{B}$ is consistent with the finding of \citet{Minato+06}.

\begin{figure}
\plotone{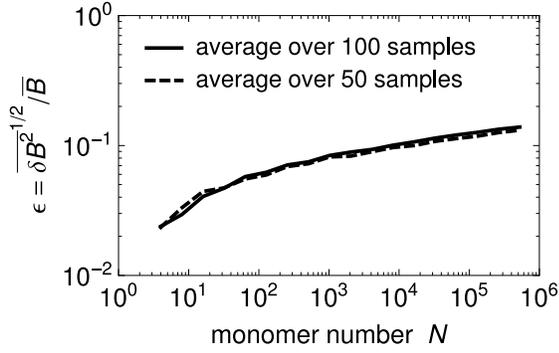}
\caption{
Normalized area dispersion $\eps = \ovl{\delta B^2}^{1/2}/\ovl{B}$ for sample BCCA clusters. 
The solid and dashed curves are obtained by averaging 100 and 50 samples, respectively.
}
\label{fig:eps}
\end{figure}

Figure~\ref{fig:eps} shows the ratio $\eps(N)$ obtained from 100 samples.
For $10 \la N \la 10^6$, $\eps(N)$ is of an order of $10^{-1}$ and increases very slowly with $N$.
Therefore, $\eps(N)$ can be well approximated as a constant $10^{-1}$.
To check the convergence, we compute $\eps(N)$ using 50 of the samples. 
The small difference between the two curves means that
the statistical error due to the finite number of samples is negligible. 

Figure~\ref{fig:eps} implies that 
$\eps(N)$ may be considerably larger than $10^{-1}$ for $N \gg 10^6$.
However, it should be noted that the above clusters has been formed 
through collisions between identical clusters.
In reality, an aggregate in an ensemble collides with aggregates of various sizes.
The most probable are collisions between aggregates of very different $B$,
since the collision probability is proportional to $|B_1-B_2|$. 
This effect generally cause the decrease in $\ovl{\delta B^2}$, and hence the decrease in $\eps$.
Therefore, the value of $\eps$ estimated here should be regarded as the upper limit of the actual values.



\end{document}